\documentclass[twocolumn]{aastex63}

\received{June 1, 2019}
\revised{January 10, 2019}
\accepted{\today}
\submitjournal{AJ}

\shorttitle{CCO II}
\shortauthors{Igoshev et al.}
\graphicspath{{./}{figures/}}

\begin{document}

\title{3D Magneto-thermal Simulations of Tangled Crustal Magnetic Field in Central Compact Objects}

\correspondingauthor{Andrei P. Igoshev}
\email{a.igoshev@leeds.ac.uk, ignotur@gmail.com}

\author[0000-0003-2145-1022]{Andrei P. Igoshev}
\affiliation{Department of Applied Mathematics, University of Leeds, LS2 9JT Leeds, UK}

\author[0000-0002-1659-1250]{Konstantinos N. Gourgouliatos}
\affiliation{Department of Physics, University of Patras, 26504 Patras, Greece}

\author[0000-0001-8639-0967]{Rainer Hollerbach}
\affiliation{Department of Applied Mathematics, University of Leeds, LS2 9JT Leeds, UK}

\author[0000-0003-1044-170X]{Toby S.~Wood}
\affiliation{School of Mathematics, Statistics and Physics, Newcastle University, Newcastle upon Tyne, NE1 7RU, UK}



\begin{abstract}
Central compact objects are young neutron stars emitting thermal X-rays with bolometric luminosities $L_X$ in the range $10^{32}$~--~$10^{34}$~erg/s.
Gourgouliatos, Hollerbach and Igoshev recently suggested that peculiar emission properties of central compact objects can be explained by tangled magnetic field configurations formed in a stochastic dynamo during the proto-neutron star stage. In this case the magnetic field consists of multiple small-scale components with negligible contribution of global dipolar field. 
We study numerically three-dimensional magneto-thermal evolution of tangled crustal magnetic fields in neutron stars. We find that all configurations produce complicated surface thermal patterns which consist of multiple small hot regions located at significant separations from each other.
The configurations with initial magnetic energy of $2.5-10\times 10^{47}$~erg have temperatures of hot regions that reach $\approx 0.2$~keV, to be compared with the bulk temperature of $\approx 0.1$~keV in our simulations with no cooling. A factor of two in temperature is also seen in observations of central compact objects. The hot spots produce periodic modulations in light curve with typical amplitudes of $\leq 9-11$~\%. Therefore, the tangled magnetic field configuration can explain thermal emission properties of some central compact objects.

\end{abstract}



\section{Introduction} \label{sec:intro}

Central compact objects (CCOs) are neutron stars (NSs) found in close proximity to the geometric center of supernova remnants (SNRs) \citep{Pavlov2004, deLuca2008, deLuca2017}.  
CCOs 
have soft X-ray luminosities of $10^{32}-10^{34}$~erg/s, incompatible with cooling luminosities of young NSs with ages of a few kyr.
Their 
X-ray spectra generally
contain 
two black-body components:
the first arising from 
bulk NS thermal emission
with temperature $\approx$0.1-0.2~keV,
and the second produced by a small, thermally emitting area (a few percent of the NS size) with temperatures of 0.2-0.4~keV.
CCO luminosities are comparable with those of some quiescent magnetars (for a review see e.g.\ \citealt{magnetar_review}).
However, CCOs show no activity similar to magnetars (with the exception of 1E 161348-5055 in SNR RCW 103, \citealt{Rea2016}, which might be a magnetar with unusual formation path). There is no indication that they are binaries. CCOs are numerous: up to 30\% of NSs associated with SNRs at distances up to 5 kps are CCOs \citep{deLuca2008},
which is difficult 
to reconcile with the Galactic NS birthrate \citep{keanekramer2008}. It is unclear how CCOs age. CCO-like objects are not found among nearby cooling NSs \citep{turolla2009} with ages of $>100$~kyr. \cite{Bogdanov2014} and \cite{Luo2015} searched for radio pulsars which show CCO-like X-ray emission and found none.

Long-term X-ray observations have revealed the spin period and period derivative for three CCOs \citep{halpern2010, gotthelf2013}.
The spin periods are: RX J0822-4300 in SNR Puppis A has $P = 112$~ms \citep{gotthelf2009}, CXOU J185238.6+004020 in Kes 79 has $P = 105$~ms \citep{gotthelf2005} and 1E 1207.4-5209 in PKS 1209-51/52 has $P = 424$~ms \citep{zavlin2000}.
From these measurements and their period derivatives,
the magnetic dipole inferred for these CCOs
can be estimated
using the formula 
$B_p \approx 3.2\times 10^{19} \sqrt{P\dot P}$~G \citep{handbook}.
The inferred values of $B_p$ are in the range $10^{10}$~--~$10^{11}$~G, which is one or two orders of magnitude smaller than the $10^{12}$~G dipole fields of young radio pulsars. The typical spin-down luminosity $\dot E$ for CCOs is $\sim 10^{32}$~erg~s$^{-1}$, which means that magnetospheric currents arising from pulsar spin-down cannot be responsible for formation of compact hot regions at the CCOs' surfaces.

CCOs with measured periods and period derivatives are located among old radio pulsars in the $P$~--~$\dot P$ plane. While these pulsars are prominent radio sources, CCOs show no non-thermal radio emission, neither persistent nor periodic. Coherent pulsar radio emission is strongly beamed and is generated in the magnetosphere by some kind of plasma instability, or as suggested recently by non-stationary plasma discharge \citep{Philippov2020}. Even if we miss the beam of radio emission due to the pulsar orientation, the presence of a pulsar wind nebula might be expected.
Young, rapidly rotating neutron stars embedded in supernova remnants are often surrounded by a compact pulsar wind nebula \citep{Gaensler_PWN2006}. This nebula is powered by the particle wind generated by rotating, magnetised NSs. No evidence of PWN is found in relation to CCOs.

Therefore, observational properties of CCOs pose multiple questions for researchers: (1) what is the source of additional heating in these NSs, (2) why are their emitting areas so small, (3) what are the descendants of CCOs, (4) why is no radio emission detected, and (5) how different CCOs really are from magnetars.
This is a part of a larger problem with variety of different NS   classes. A grand unification of NSs is a scenario which tries to explain the difference between various classes of NSs based on properties and evolution of their magnetic fields \citep{Kaspi2010, guns2014}. In the framework of this scenario, CCOs might evolve into some other category of NSs, such as radio pulsars at timescales of $10^4$~--~$10^6$~years.

Several scenarios have been proposed to answer these questions and to explain the origin and evolution of CCOs. These scenarios attribute the additional source of energy to strong hidden magnetic fields of NSs.
Among these scenarios are fall-back accretion and a stochastic dynamo \citep{gotthelf2007}. Magnetic fields of NSs are multi-component; they can include both poloidal and toroidal parts, as well as small-scale fields. Only the large-scale  poloidal dipole magnetic field emerging from the surface can be measured using the timing technique. 

In the fall-back accretion scenario,
some material expelled by the supernova explosion does not have enough energy to escape from the system indefinitely, and is subsequently accreted back onto the NS \citep{Chevalier1989}. This accretion continues after the NS crust has solidified. New infalling material buries 
the magnetic field,
covering it with 
a new crustal layer \citep{BernalCCO}. The buried magnetic field gradually re-emerges through the surface over time \citep{Ho2011, Ho2012, BernalCCO2013}.  In this scenario, the excessive heating is explained by the amplification and dissipation of the magnetic field,
which is compressed 
between the superconducting NS core and new crust formed from fall-back material.
The absence of 
radio emission is explained by the suppression of the surface dipolar magnetic field, preventing the NS from producing enough plasma in its magnetosphere. The same explanation holds for the lack of a pulsar wind nebula. 
\cite{igoshev2016} studied the evolution of small-scale magnetic field after a fall-back and found that the field (both small- and large-scale) re-emerges, so these NSs should start operating as normal radio pulsars after $10^4$~--~$10^5$~years. It is still unclear if the heat produced by the magnetic field causes a formation of small emitting regions at the surface in this scenario. If a strong fall-back occurs at the newly-born magnetar, a CCO-magnetar could be produced. This is a CCO which has an extremely strong magnetic field trapped deep in the crust which should occasionally exhibit flares.

In the stochastic dynamo scenario,
the NS is born with a predominantly small-scale magnetic field, with only a weak dipole component.
This small-scale field is formed during
the proto-NS stage, which lasts for tens of seconds \citep{pons1999} and ends with the solidification of the crust. During this stage there are two regions where the matter is unstable according to the Ledoux convection criterion. One of these regions is located around densities of $10^{13}$~g~cm$^{-3}$ of proto-NSs \citep{thompson2001}. \cite{nagakura2020} analysed
state-of-the-art 
3D simulations of supernova explosions for progenitor masses ranging from $9$~M$_\odot$ to $20$~M$_\odot$ and found that convection always develops in proto-NSs. This convection zone is an ideal place for a dynamo which could generate magnetic fields in the range $10^{12}$~--~$10^{15}$~G.

 The generation of a large-scale magnetic field is thought to require rapid rotation of the proto-NS (see e.g.\ recent simulations by \citealt{raynaud2020}). If the proto-NS rotates slowly, a dynamo could still operate, but producing only a strong small-scale field \citep{thompson2001}. Such a field with surface $B_s \approx 10^{14}$~G is hidden for a distant observer, because it hardly contributes to the NS spin-down. A stochastic dynamo could also operate in convective fall-back flow \citep{thompson2001}, and also lead to formation of small-scale magnetic fields.

In the stochastic dynamo scenario, the excessive heating of CCOs could be explained by Ohmic decay combined with fast Hall evolution of this much stronger non-dipolar magnetic field. We test it in our research using the tangled configuration of magnetic field suggested by \cite{CCOI}.
In these configurations, the magnetic fields are comparable in strength to magnetar ones, but should not produce similar stresses and therefore, should not lead to magnetar-like activity.
\cite{CCOI,branbedburg2020} found that these configurations where the poloidal dipole is absent or rather small develop such a component in the course of their evolution due to an inverse Hall cascade.

The Hall cascade was first suggested for electron-MHD by \cite{Goldreich1992} with application to NS crusts. Occurrence of this cascade was confirmed in detailed numerical simulations (see e.g.\ \citealt{wareing2009,wareing2010,branbedburg2020}). Turbulence leads to redistribution of magnetic energy to both small-scale (forward cascade) and large-scale (inverse cascade) fields. In the case of tangled magnetic field configurations, an inverse cascade leads to an increase of the dipolar magnetic field with time. The small dipolar component at the beginning of NS evolution is not sufficient for the activation of the radio-pulsar mode. When the global dipolar component rises after $\sim 10$~kyr of evolution, this object might start operating as a radio pulsar, which solves the problem regarding the absence of CCO descendants. Because in this case CCOs turn into normal radio pulsars with no unusual properties after a period of time.  

This article is structured as follows. In Section~\ref{s:method} we describe the method which we use in our simulations; in Section~\ref{s:res_mhd} we present results of three-dimensional simulations and corresponding light curves. 
We discuss all results and conclude in Sections~\ref{s:discuss} and \ref{s:conclusion} respectively.

\section{Methods}
\label{s:method}
We perform simulations in two steps. First we compute the surface temperature distribution using an updated version of the MHD code \texttt{PARODY} \citep{Dormy1998, wood2015,gourgouliatos2016}. Next we compute the light curves and pulsed fraction for different orientations of rotating NSs using the ray-tracing in general relativity.

\subsection{Magneto-thermal evolution}
The details of the magneto-thermal code are summarised in \cite{grandis2020, igoshev2020}. We solve magnetic induction and thermal diffusion equations in the solid crust of NS using the electron-MHD approximation in which only electrons carry electric charge and thermal flux. 
Basically, we solve two coupled partial differential equations. The first is the induction equation:
$$
\frac{\partial \vec B}{\partial t} = - c\vec \nabla \times \left\{\frac{1}{4\pi e n_e}(\vec \nabla \times \vec B)\times \vec B + \frac{c}{4\pi\sigma} \vec \nabla \times\vec B\right. 
$$
\begin{equation}
\left.\hspace{4.5cm}- \frac{1}{e} S_e \nabla T\right\}
\label{e:ind}
\end{equation}
where $\vec B$ is the magnetic field, $n_e$ is the electron number density, $e$ is the elementary charge, $c$ is the speed of light, $\sigma$ is the electric conductivity, $S_e$ is the electron entropy, and $T$ is the temperature.
The second is the heat 
equation:
$$
C_V \frac{\partial T}{\partial t} = \vec \nabla \cdot (\hat k \cdot \vec \nabla T) + \frac{|\vec \nabla \times B|^2 c^2}{16\pi^2 \sigma} \hspace{2cm}
$$
\begin{equation}
\hspace{4cm}+ \left(\frac{c}{4\pi e}\right) T\vec \nabla S_e \cdot (\vec \nabla \times \vec B)
\label{e:diff}
\end{equation}
where $\hat k$ is the thermal conductivity tensor. 
The first two terms in our induction eq.~(\ref{e:ind}) are the ones introduced by \cite{Goldreich1992}, and represent the Hall evolution and Ohmic dissipation, respectively. The third term describes the so-called Biermann battery effect, i.e.~magnetic field generation due to temperature gradients \citep{Blandford:1983}.
This effect is closely related to electron baroclinicity, which acts as a source of electron circulation, and hence magnetic induction.

In the heat 
eq.~(\ref{e:diff}), the first term represents anisotropic thermal diffusion. While heat flows freely along magnetic field lines, heat transfer is inhibited in the direction orthogonal to field lines. The second term represents heating by Ohmic field decay, and the third term represents the entropy carried by electric currents. Unless strong temperature gradients are maintained externally, e.g.~by magnetospheric heating \citep{grandis2020}, the last terms in eqs.~(\ref{e:ind},\ref{e:diff}) are generally very small in comparison to other terms, and play a negligible role in the field evolution.

We write the thermal conductivity tensor in the following form:
\begin{equation}
(\hat k^{-1})_{ij} = \frac{3e^2}{\pi^2 k_B^2 T}\left(\frac{1}{\sigma} \delta_{ij} + \frac{\epsilon_{ijk} B_k}{ecn_e}\right)    
\end{equation}
where $\delta_{ij}$ is the Kronecker delta, $\epsilon_{ijk}$ is the Levi-Civita symbol, $k_B$ is the Boltzmann constant, and $B_k$ is the field component along the $k$-axis.

Electron entropy is related to temperature and chemical potential $\mu(r)$ of a degenerate, relativistic Fermi gas as follows:
\begin{equation}
S_e = \frac{\pi^2k_B^2 T}{\mu}    
\end{equation}

Similar to \citet{gourgouliatos14b} we use the following radial profiles for the chemical potential, electron density and conductivity:
\begin{eqnarray}
  \mu(r) &=&\mu_0 \left(1 + \frac{1-r}{0.0463}\right)^{4/3} \\
  n_e(r) &=& n_0 \left(1 + \frac{1 - r}{0.0463} \right)^{4} \\
  \sigma(r) &=& \sigma_0 \left(1 + \frac{1 - r}{0.0463} \right)^{8/3}
\end{eqnarray}
where $\mu_0 = 2.9\times 10^{-5}$~erg, $n_0 = 2.5\times10^{34}$cm$^{-3}$ and $\sigma_0 = 1.8\times10^{23}$s$^{-1}$.
For computational convenience
we assume that the heat capacity is proportional to the temperature in the form (see e.g.\ \citealt{Page2004})
\begin{equation}
C_V \propto \sigma T    
\end{equation}
so that eq.~(\ref{e:diff}) becomes a linear equation in $T^2$.

\subsection{Boundary and initial conditions}

For the magnetic induction at the inner boundary, we assume the Meissner condition, i.e. zero radial component of magnetic field \ $B_r = 0$ and zero tangential component of the current $J_t = 0$  at the crust-core boundary, see \cite{grandis2020} and \cite{hollerbach2004} for more details. Physically this corresponds to a scenario in which the field is expelled from the core before the crust solidifies. In reality, the core might still contain some magnetic field,
although the timescale on which the core field evolves is uncertain,
see e.g.\ \cite{Gusakov2020}. For the magnetic field at the outer boundary
we use vacuum boundary conditions,
i.e.~we assume that
$\vec \nabla \times \vec B = 0$
in the region outside the star.
This condition is implemented numerically as:
\begin{equation}
\frac{\partial B_{\mathrm{p}, l}^m}{\partial r} + \frac{l+1}{r} B_{\mathrm{p}, l}^m = 0 
\end{equation}
and
\begin{equation}
B_{\mathrm{t}, l}^m = 0    
\end{equation}
where $B_{\mathrm{p}, l}^m$ and $B_{\mathrm{t}, l}^m$
represent the coefficients of the poloidal and toroidal magnetic potentials,
which are expanded in series of spherical harmonics
of degree $l$ and order $m$.
Physically these boundary conditions correspond to neglecting any electric currents in the star's tenuous plasma atmosphere.
We note that CCOs have no radio emission, and no pulsar wind nebula was ever detected around a CCO, so their magnetospheres might have less plasma than is typical for normal radio pulsars.

For the temperature, at the crust core interface we assume no cooling, yielding the boundary condition $\partial T/\partial t = 0$.
In reality the core cools at a rate determined by neutrino emission, which is practically insensitive to conditions within the crust.
Since our focus in this study is on small-scale temperature anomalies produced within the crust
we neglect the core cooling for simplicity.
At the outer boundary we assume that
the heat flux out of the computational domain
is subsequently emitted as black-body radiation from the star's surface, i.e.
\begin{equation}
-\vec r \cdot \hat \kappa \cdot \nabla T|_b = \sigma_S T_s^4    
\end{equation}
where $\sigma_S$ is the Stefan-Boltzmann constant,
$T_b$ is the temperature at the top of the computational domain, and $T_s$ is the effective surface temperature.
We assume that
these two temperatures
are related by
the thermal blanket equation
\begin{equation}
\left(\frac{T_b}{10^8\; \mathrm{K}} \right) = \left(\frac{T_s}{10^6\; \mathrm{K}}\right)^2,
\label{e:temp_s}
\end{equation}
similar to a relation suggested by \cite{Gudmundsson1983}, but in a more simplified form. This assumption is necessary in our calculations because the thermal relaxation timescale of the envelope is much shorter than any other evolution timescale involved in our simulations. As was shown by \cite{Gudmundsson1982}, the effective surface temperature depends essentially only on the temperature in deep layers (below densities $10^{10}$~g~cm$^{-3}$) and surface gravity. 

We use the same magnetic field configurations as in \cite{CCOI}, so the initial magnetic field consists of several high-order multipoles with $10\leq l \leq 20$. We choose phases randomly for these fields. The amount of initial magnetic energy $E_\mathrm{tot,0}$ in this field is set at the beginning of simulations.
A dipolar poloidal magnetic field is added to the stochastic magnetic field configuration with fixed $B_\mathrm{dip, 0}$ value before the simulations are started.

We compute the same first five models as in \cite{CCOI}. In models 6 and 7 we decrease the total magnetic energy in the crust by a factor of 4. We make this change because for larger energies we encounter certain numerical problems. We summarise the details of our simulations in Table~\ref{t:models}.

\begin{table}[]
    \centering
    \begin{tabular}{cccccccc}
    \hline
    Name     &  $E_\mathrm{tot,0}$ & $B_\mathrm{dip, 0}$ & $\langle B_0 \rangle$ & $p_\mathrm{max,1}$ & PF$_\mathrm{m}$ \\
             &   (erg)             &    (G)              &  (G)                  & &  \%\\
    \hline
    Model 1  & $2.5\times 10^{45}$ & 0                   & $2\times 10^{14}$     & 0.58 & 0.3 \\
    Model 2  & $2.5\times 10^{45}$ & $10^{10}$           & $2\times 10^{14}$     & 0.58 & 0.3 \\
    Model 3  & $2.5\times 10^{45}$ & $10^{11}$           & $2\times 10^{14}$     & 0.58 & 0.3 \\
    Model 4  & $2.5\times 10^{47}$ & $10^{10}$           & $2\times 10^{15}$     & 0.44 & 11\\
    Model 5  & $2.5\times 10^{47}$ & $10^{12}$           & $2\times 10^{15}$     & 0.43 & 11\\
    Model 6  & $1\times 10^{48}$ & $10^{10}$             & $5\times 10^{15}$     & 0.39 & 9\\
    Model 7  & $1\times 10^{48}$ & $10^{12}$             & $5\times 10^{15}$     & 0.39 & 9\\
    \hline     
    \end{tabular}
    \caption{Numerical models computed here. $p_\mathrm{max,1}$ is the value of total magnetic energy decay exponent at its first maximum. PF$_\mathrm{m}$ is the maximum pulsed fraction computed for all models at age 3.5~kyr. }
    \label{t:models}
\end{table}

\subsection{Light curves and pulsed fraction}
In order to compute the light curves we follow the same technique as described in \cite{igoshev2020}. Namely, NS orientation is described using three angles: $i$ is the angle between line of sight and rotational axis of NS, $\kappa$ is the angle between magnetic dipole moment and rotational axis, and $\Delta \Phi$ is the initial phase. These three angles uniquely describe NS orientation at the start of rotation. To simulate rotation of NS, we use simplified equations presented by \cite{beloborodov2002} combined with:
\begin{eqnarray}
\mu =& \sin i \left[ \cos \Phi \cos \theta \sin \kappa \right. \nonumber \\
&- \sin \theta (\sin \phi \sin\Phi - \cos \kappa \cos \phi \cos \Phi) \left. \right] \nonumber \\
&+ \cos i \left[ \cos \kappa \cos \theta - \cos \phi \sin \kappa \sin \theta \right]
\label{e:mu}
\end{eqnarray}
where $\theta$ and $\phi$ are coordinates at the NS surface with respect to original dipolar magnetic poles, $\mu=\cos\psi$ is the direction toward the position with coordinates $\theta,\phi$ at infinity, and $\Phi\in [0,2\pi]$ is the angular phase. Our eq.~(\ref{e:mu}) reduces to eq.\ (5) of  \cite{beloborodov2002} if a hot spot is located at the magnetic pole, i.e.\ $\theta=0=\phi$.

We choose the photon beaming to be proportional to  $\cos^2 \theta'$ \citep{dedeo2001} because of the following argument. In an atmosphere which is not affected by magnetic field, the darkening is described by the Hopf function, see e.g.\ \cite{chandrasekhar_book}. The intensity emitted at angle $90^\circ$ is $\approx 1/3$ of the intensity emitted in the normal direction. In a strong magnetic field \citep{vanAdelsberg2006} the beaming is stronger than this, so at angles of $80^\circ$ the emission is 0.1 of its value at $10^\circ$, and it goes to 0 at $90^\circ$ (see figure 15 of \citealt{vanAdelsberg2006}), which is roughly similar to a $\cos^2 \theta'$ behaviour.
For CCOs the pulsed fractions are typically small, so it is better to overestimate the pulsed fraction in our models to check if the model has enough predictive power.

The flux emitted by each surface element of NS is integrated over the visible hemisphere (which is slightly more than half of the star due to the light bending in a strong gravitational field) to produce the visible flux for each rotational phase. We use 16 phases to simulate a light curve. 
For a few orientations we compute the pulsed fraction (PF) as:
\begin{equation}
PF = \frac{F_\mathrm{max}-F_\mathrm{min}}{F_\mathrm{max}+F_\mathrm{min}},
\end{equation}
where $F_\mathrm{max}$ and $F_\mathrm{min}$ are maximum and minimum fluxes for a fixed NS rotational orientation. We select the maximum pulsed fraction among different orientations and show them in Table~\ref{t:models}.

In \cite{igoshev2020}, we fitted observed light curves of magnetars in quiescence to estimate the NS rotational orientation. This was possible because magnetars seem to have regular large-scale magnetic fields. This is not the case for CCOs. If their magnetic fields are indeed formed as a result of a stochastic dynamo, the dipolar component is weak and changes location relatively quickly (see \citealt{CCOI}). Additional loops of magnetic field are located randomly; therefore, it is necessary to describe the location of each individual loop of magnetic field to reproduce a thermal map. Thus, there is no use in exact fitting of the light curve, since multiple slightly different magnetic field configurations will result in very similar light curves but slightly different NS rotational orientations.

\section{Results}
\label{s:res_mhd}

\begin{figure*}
\gridline{\fig{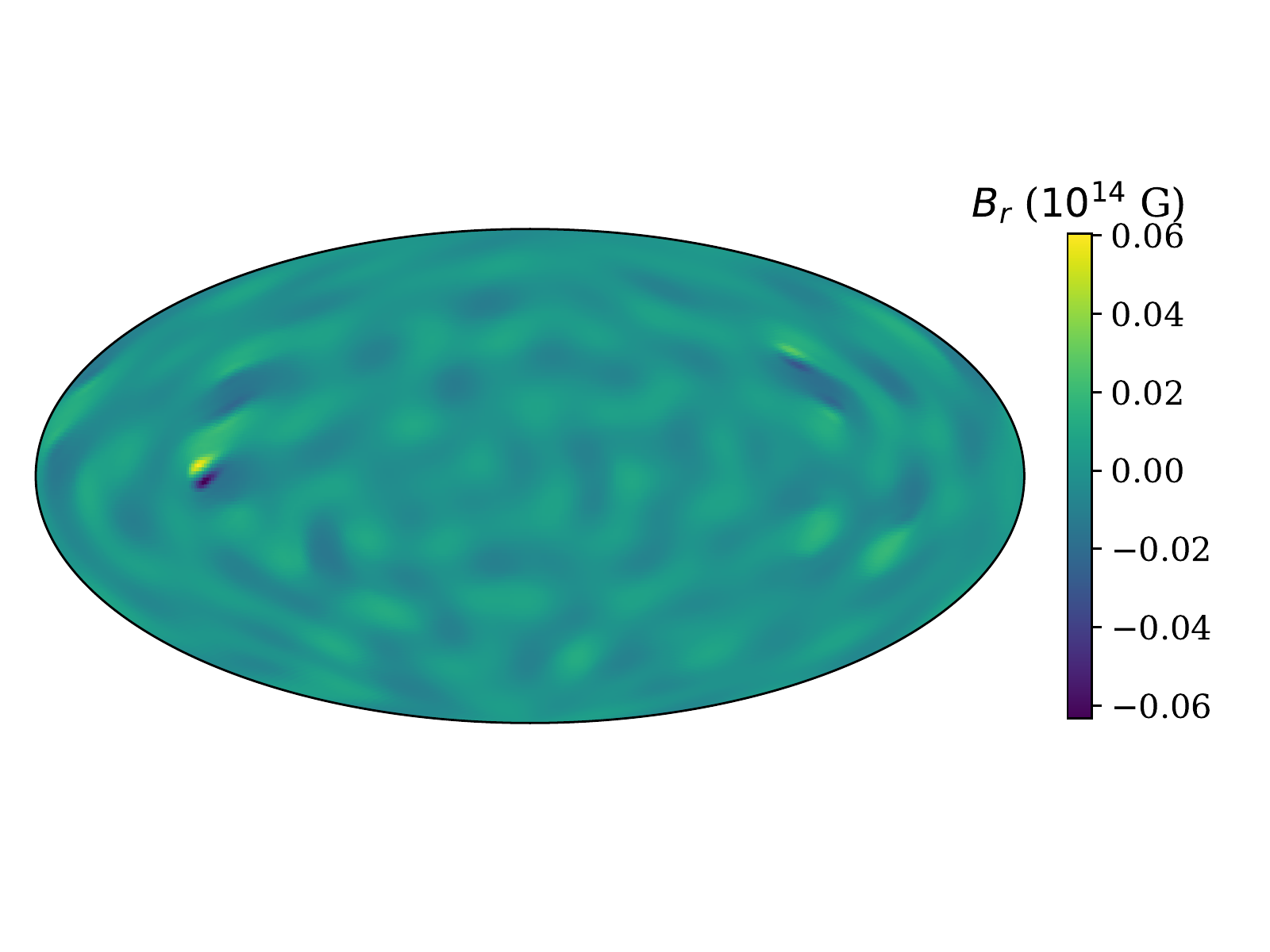}{0.3\textwidth}{(a)}
          \fig{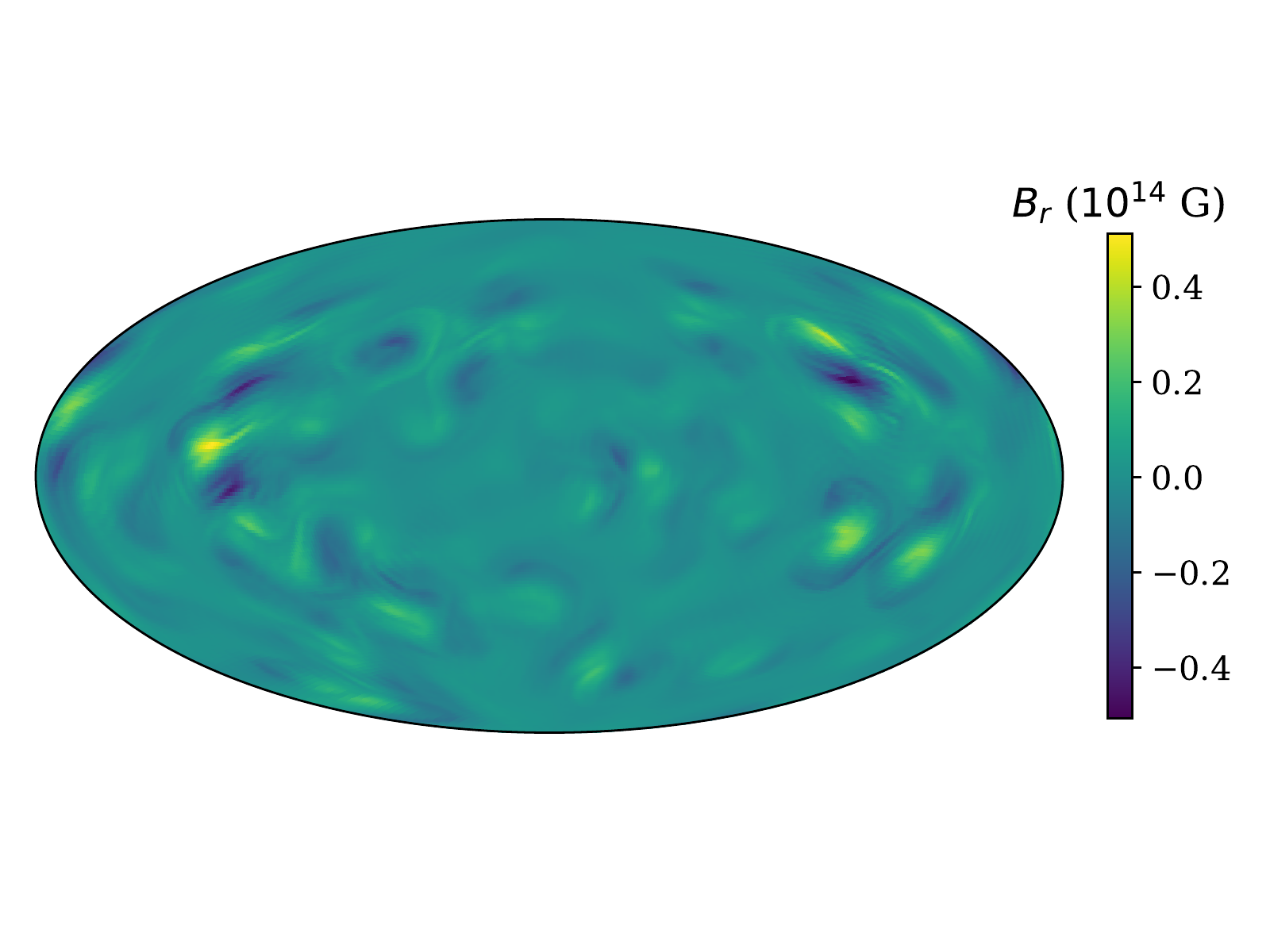}{0.3\textwidth}{(b)}
          \fig{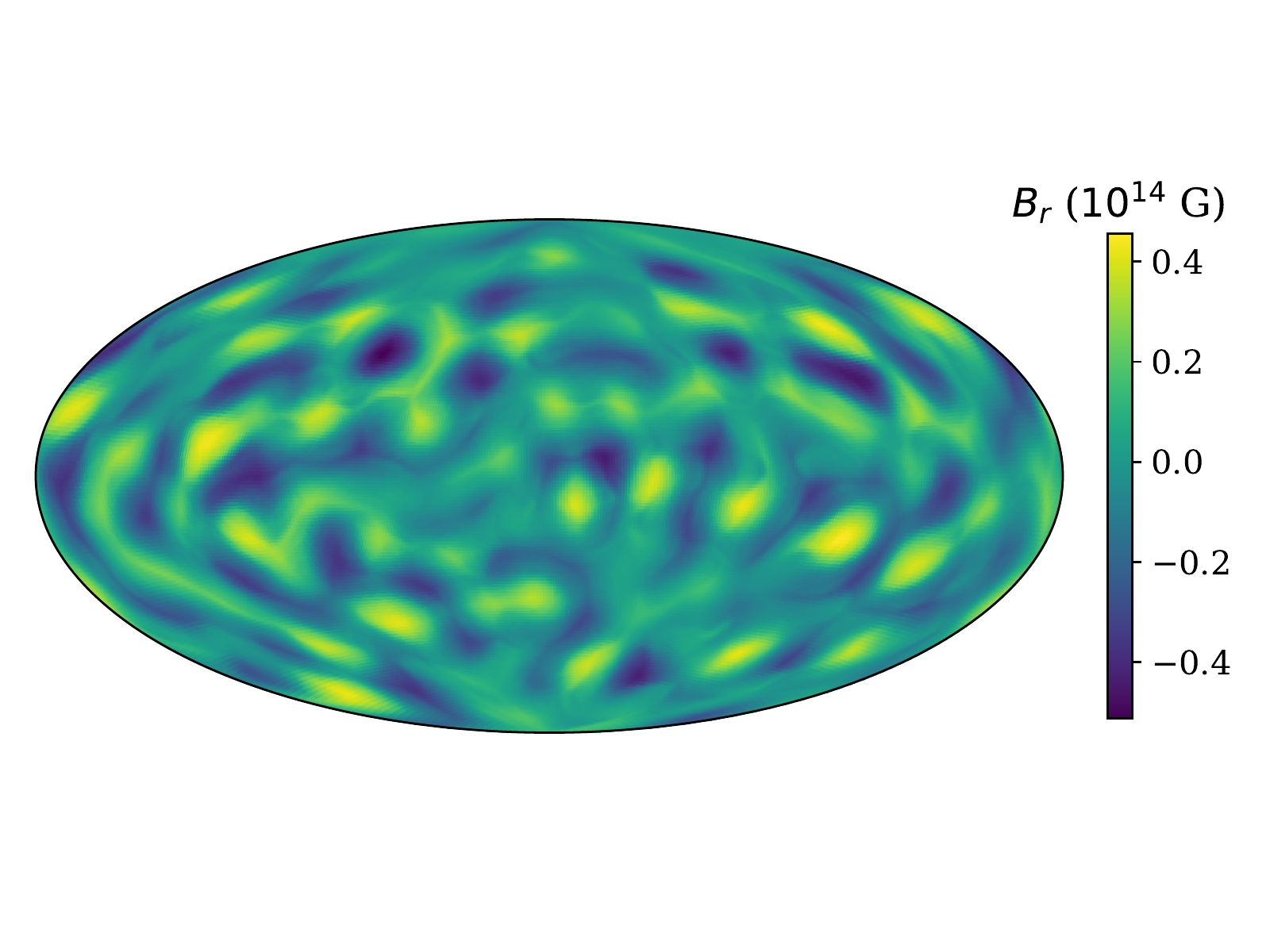}{0.3\textwidth}{(c)}
          }
\gridline{\fig{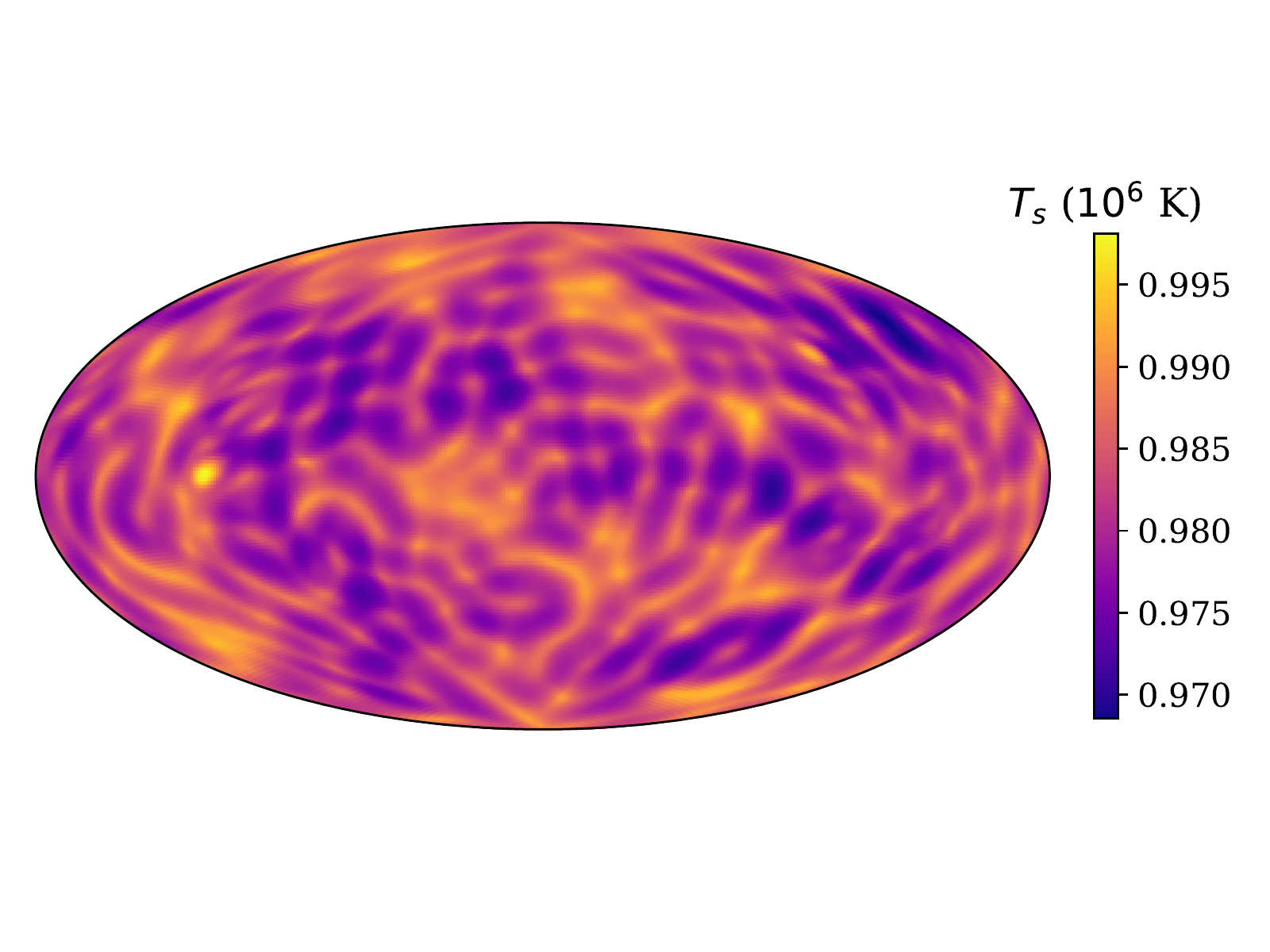}{0.3\textwidth}{(d)}
          \fig{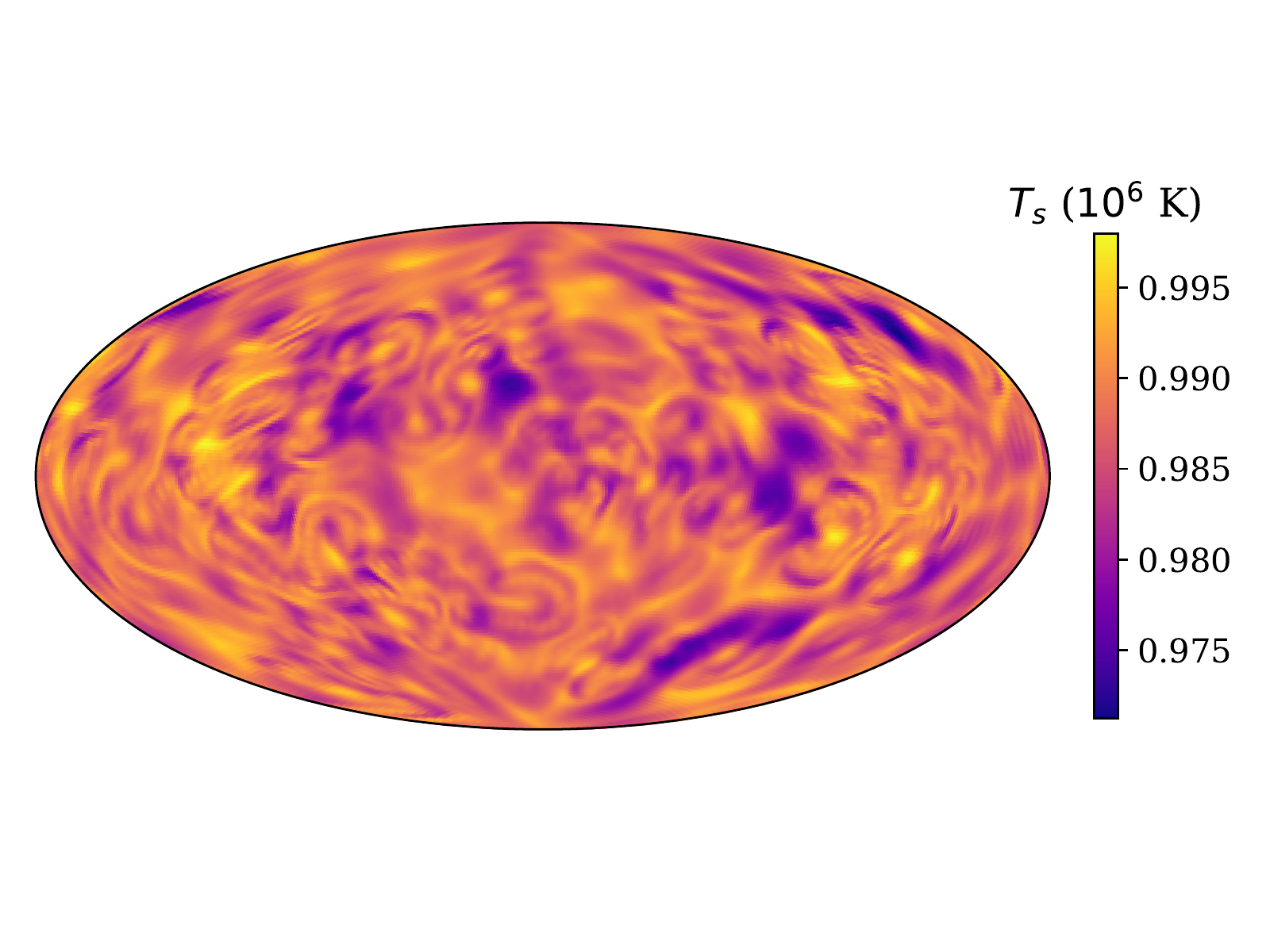}{0.3\textwidth}{(e)}
          \fig{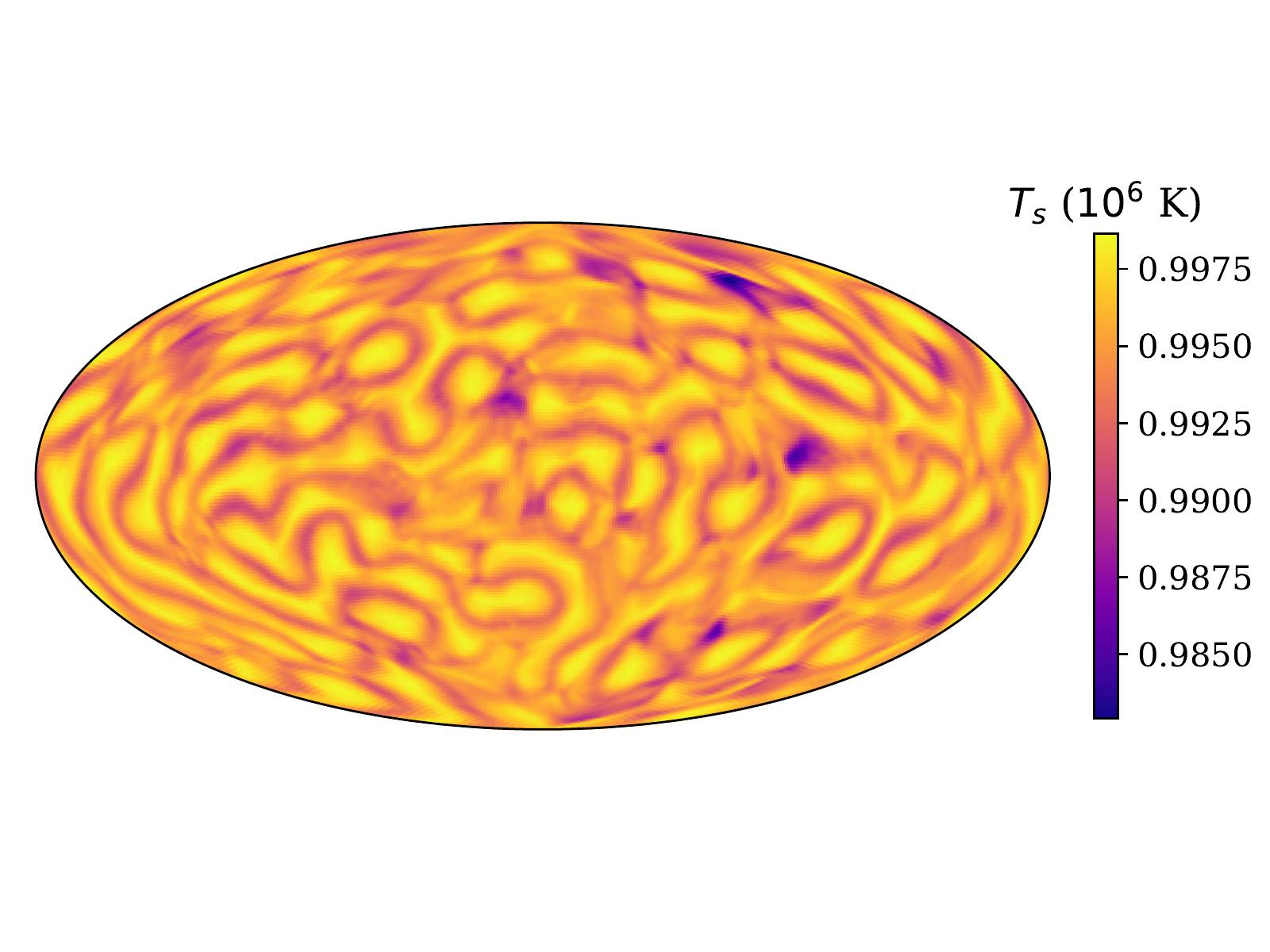}{0.3\textwidth}{(f)}
          }
\caption{Surface magnetic fields (upper row) and surface temperatures (lower row) for model 2. Individual panels correspond to different ages: (a, d) 3.5~Kyr, (b, e) 9.5~Kyr, (c,f) 37.7~Kyr.
\label{f:m2}}
\end{figure*}

\begin{figure*}
\gridline{\fig{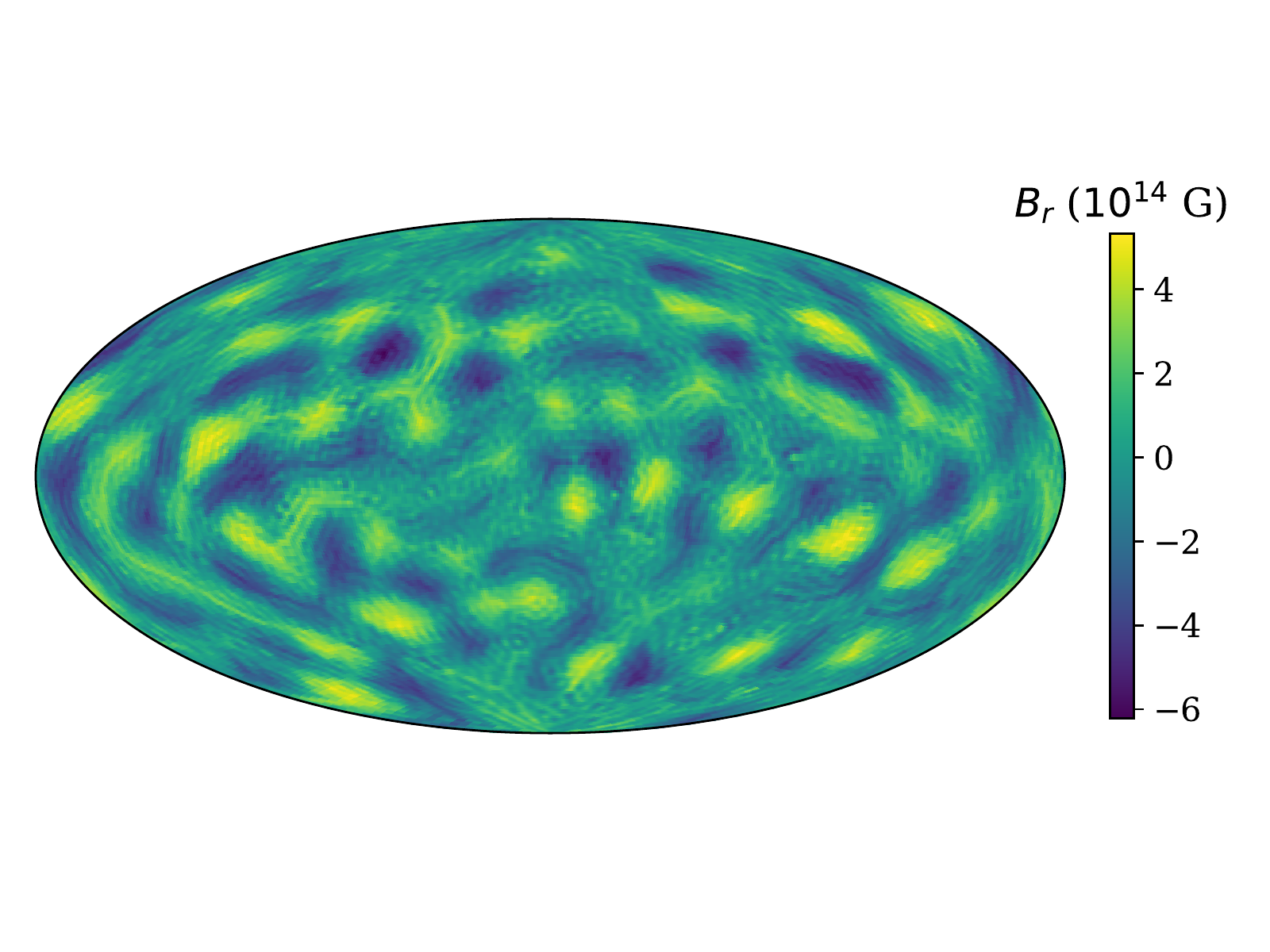}{0.3\textwidth}{(a)}
          \fig{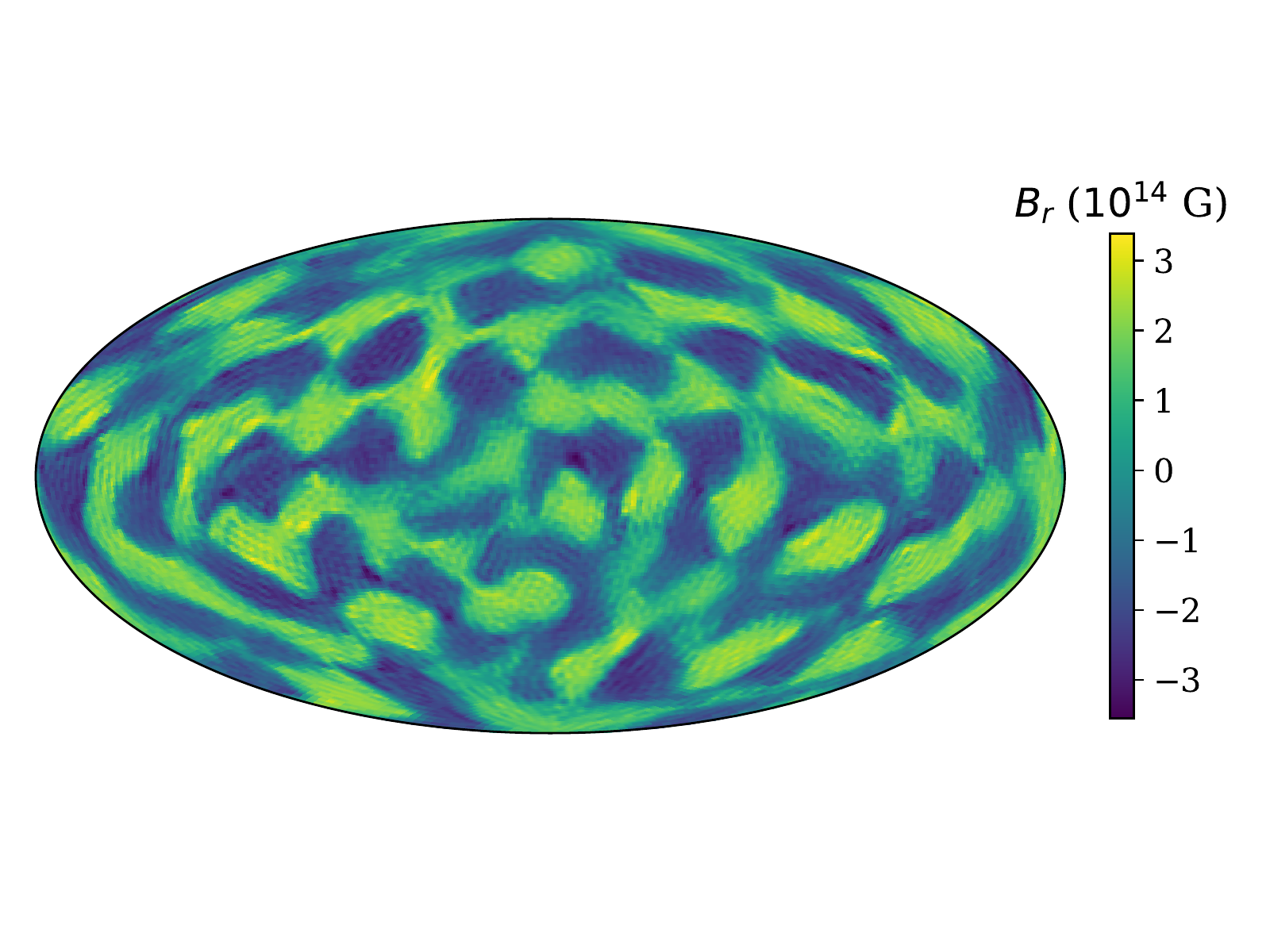}{0.3\textwidth}{(b)}
          \fig{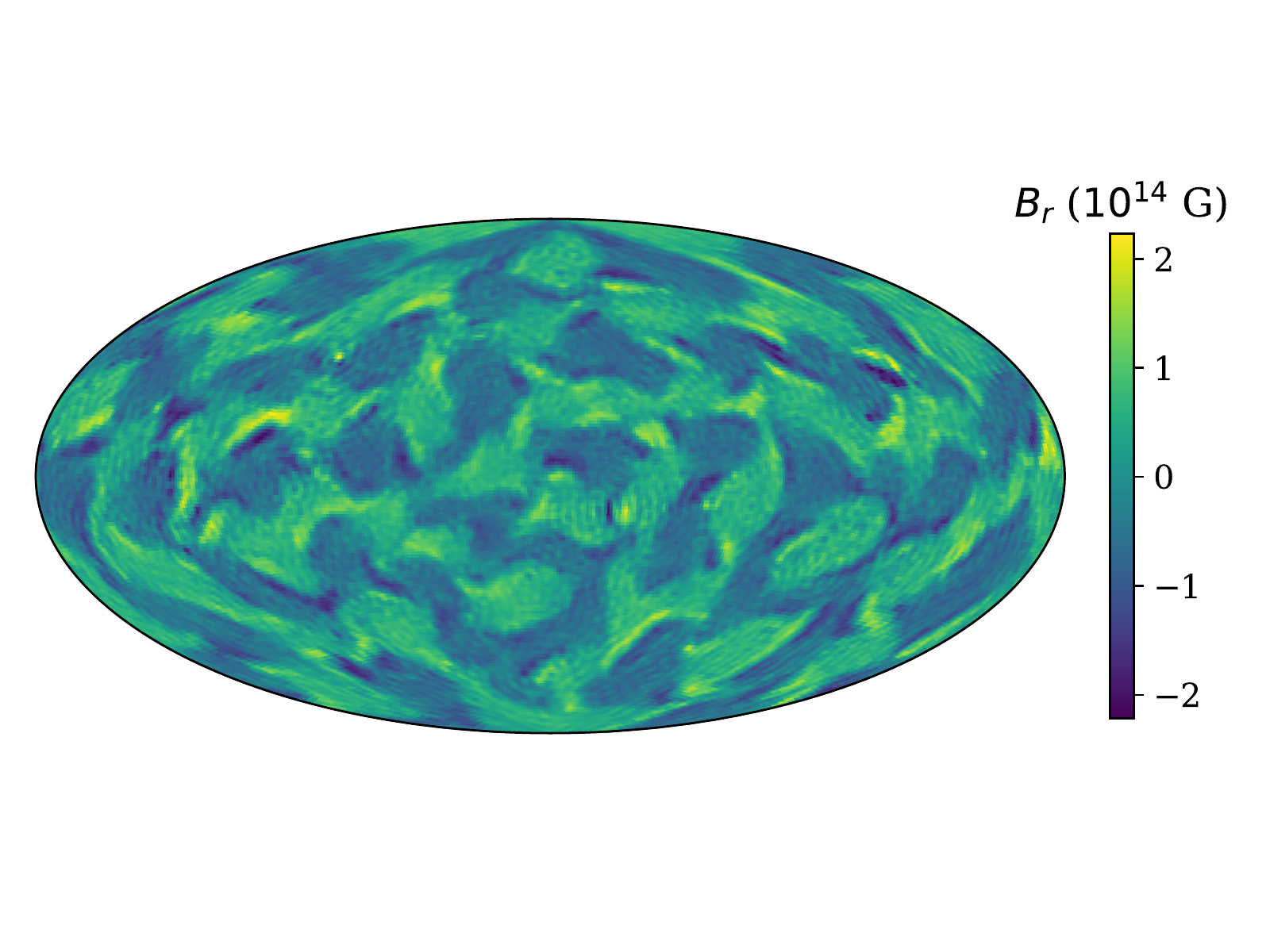}{0.3\textwidth}{(c)}
         }
\gridline{\fig{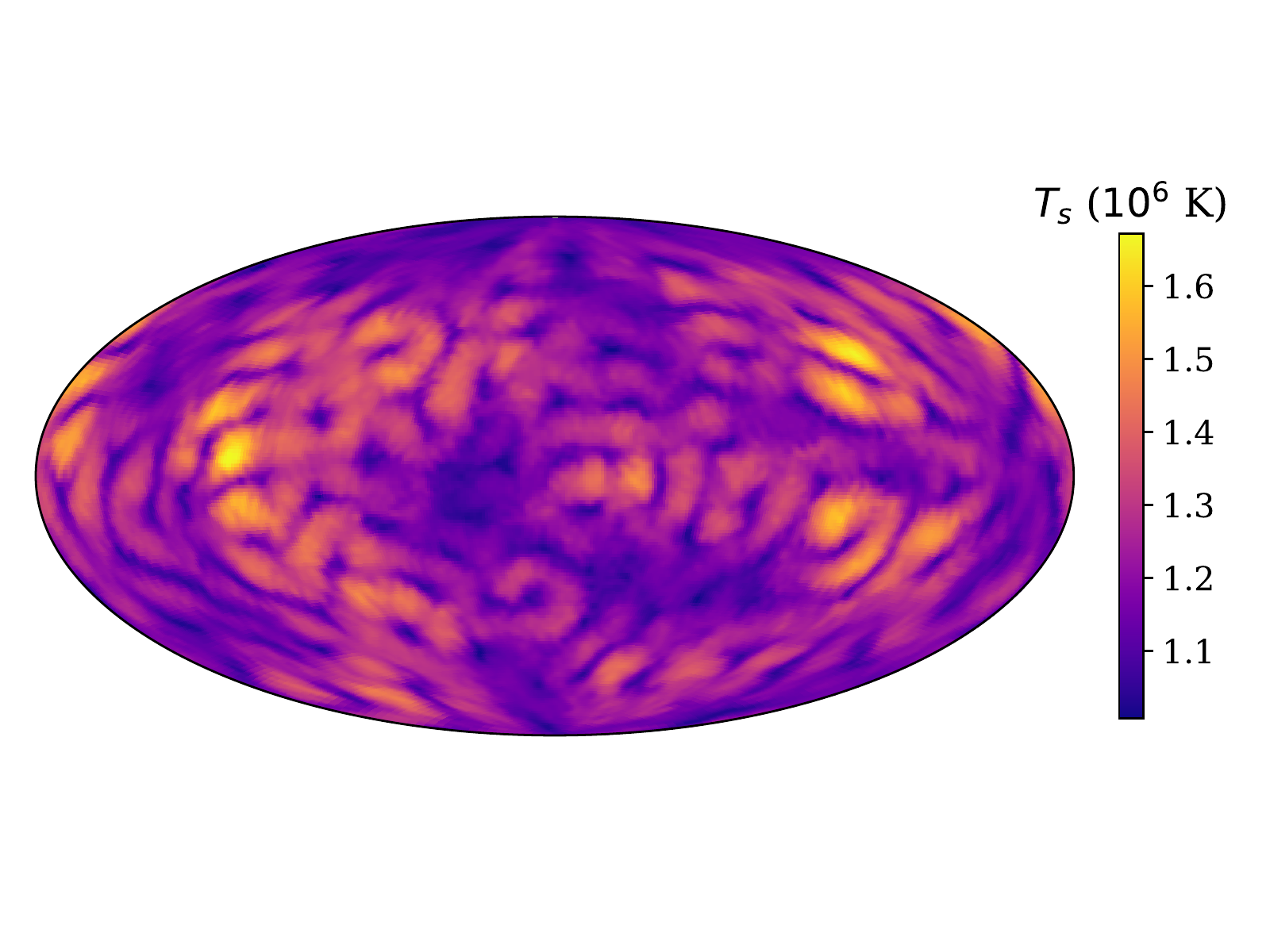}{0.3\textwidth}{(d)}
          \fig{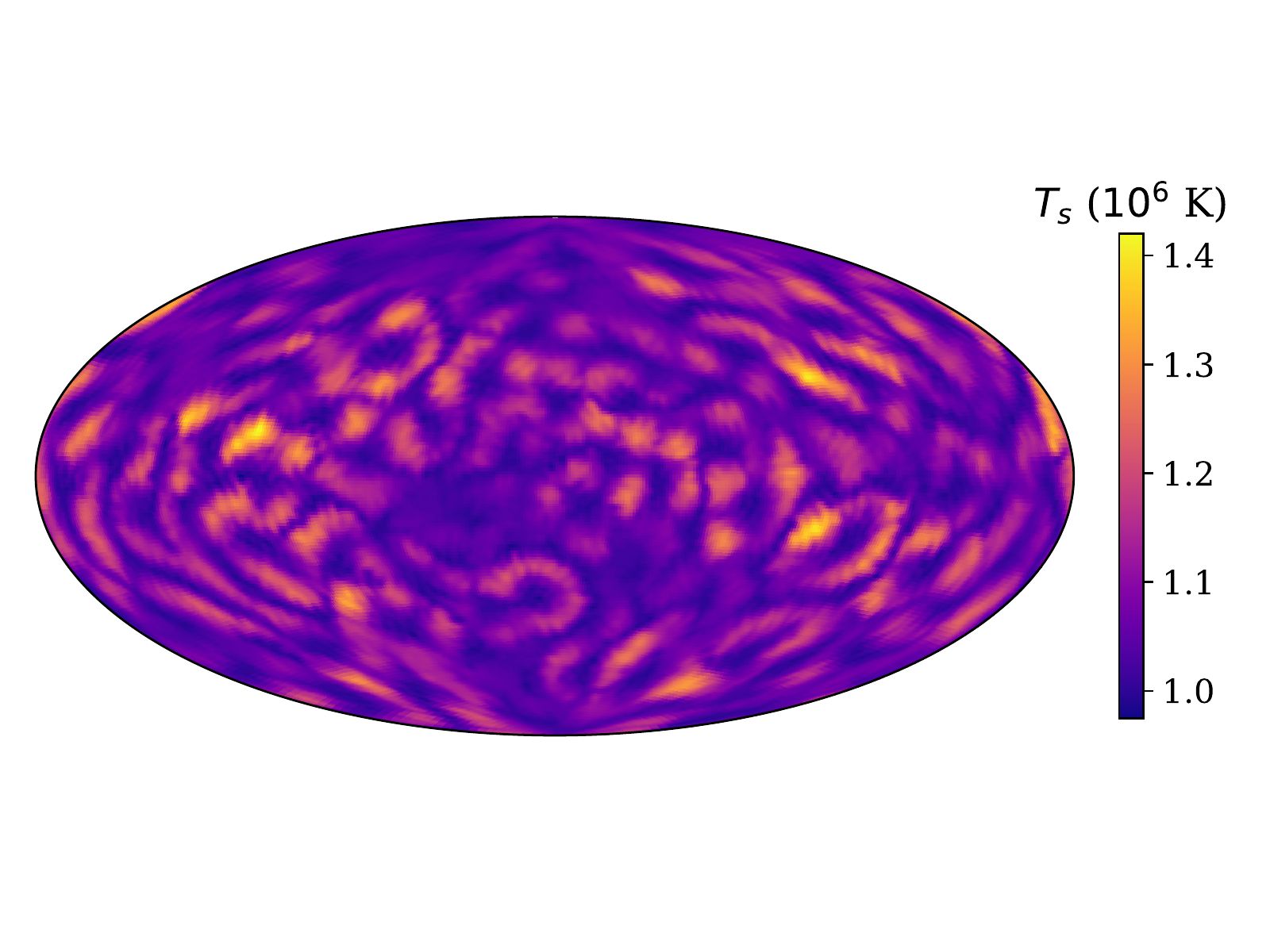}{0.3\textwidth}{(e)}
          \fig{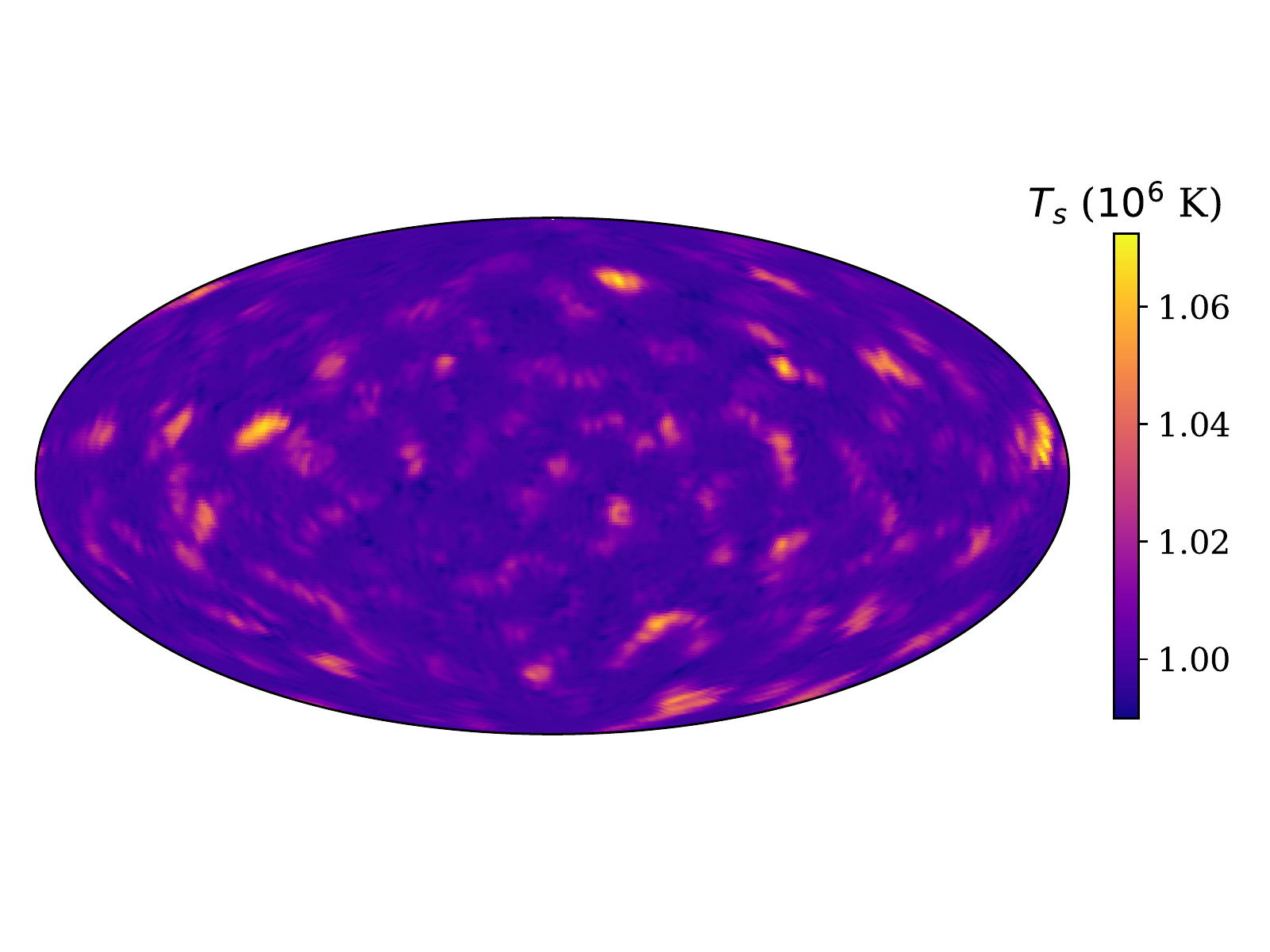}{0.3\textwidth}{(f)}
          }
\caption{Surface magnetic fields (upper row) and surface temperatures (lower row) for model 4. Individual panels correspond to different ages: (a, d) 3.5~Kyr, (b, e) 9.5~Kyr, (c,f) 37.7~Kyr.
\label{f:m4}}
\end{figure*}

\begin{figure*}
\gridline{\fig{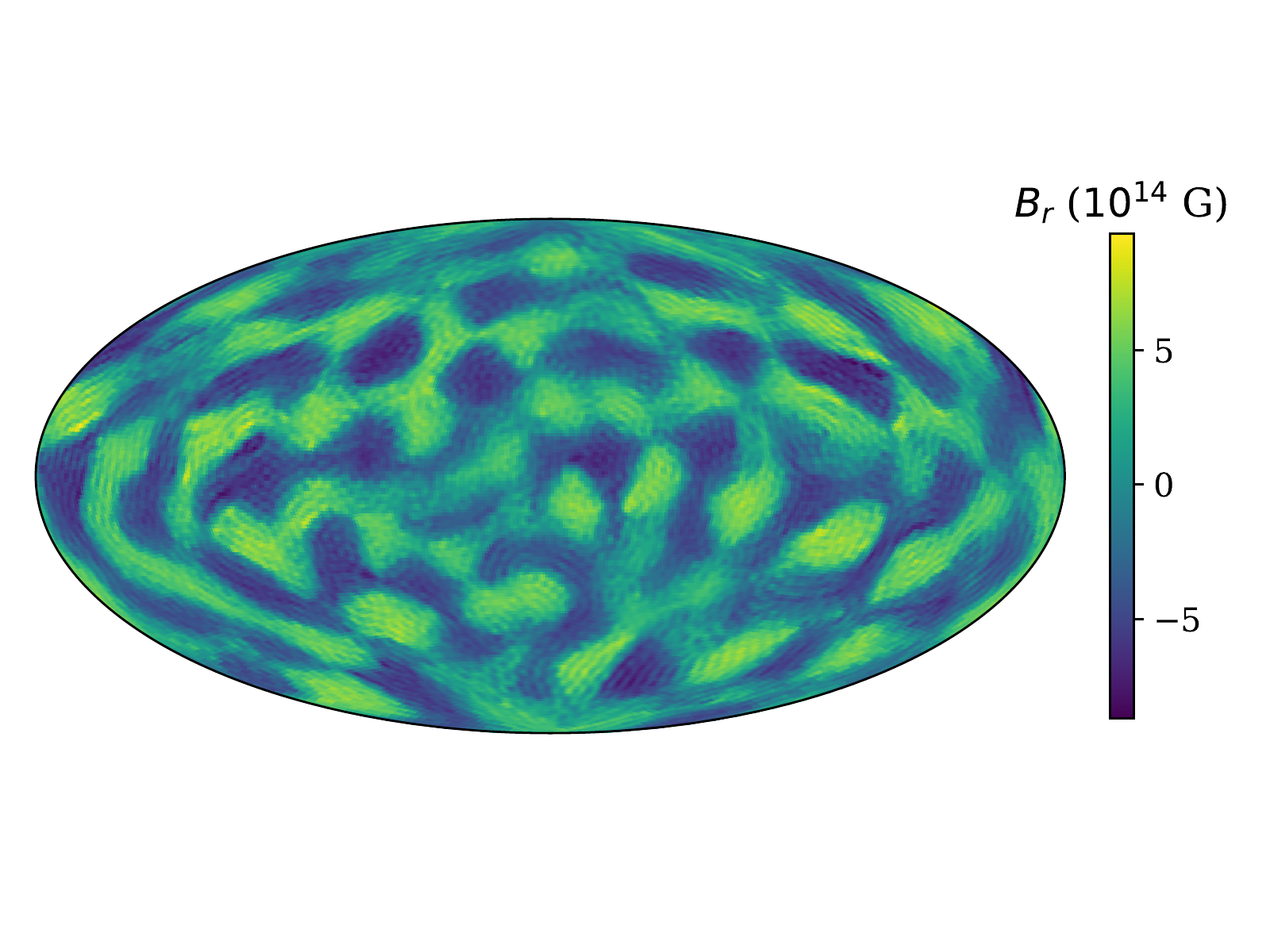}{0.3\textwidth}{(a)}
          \fig{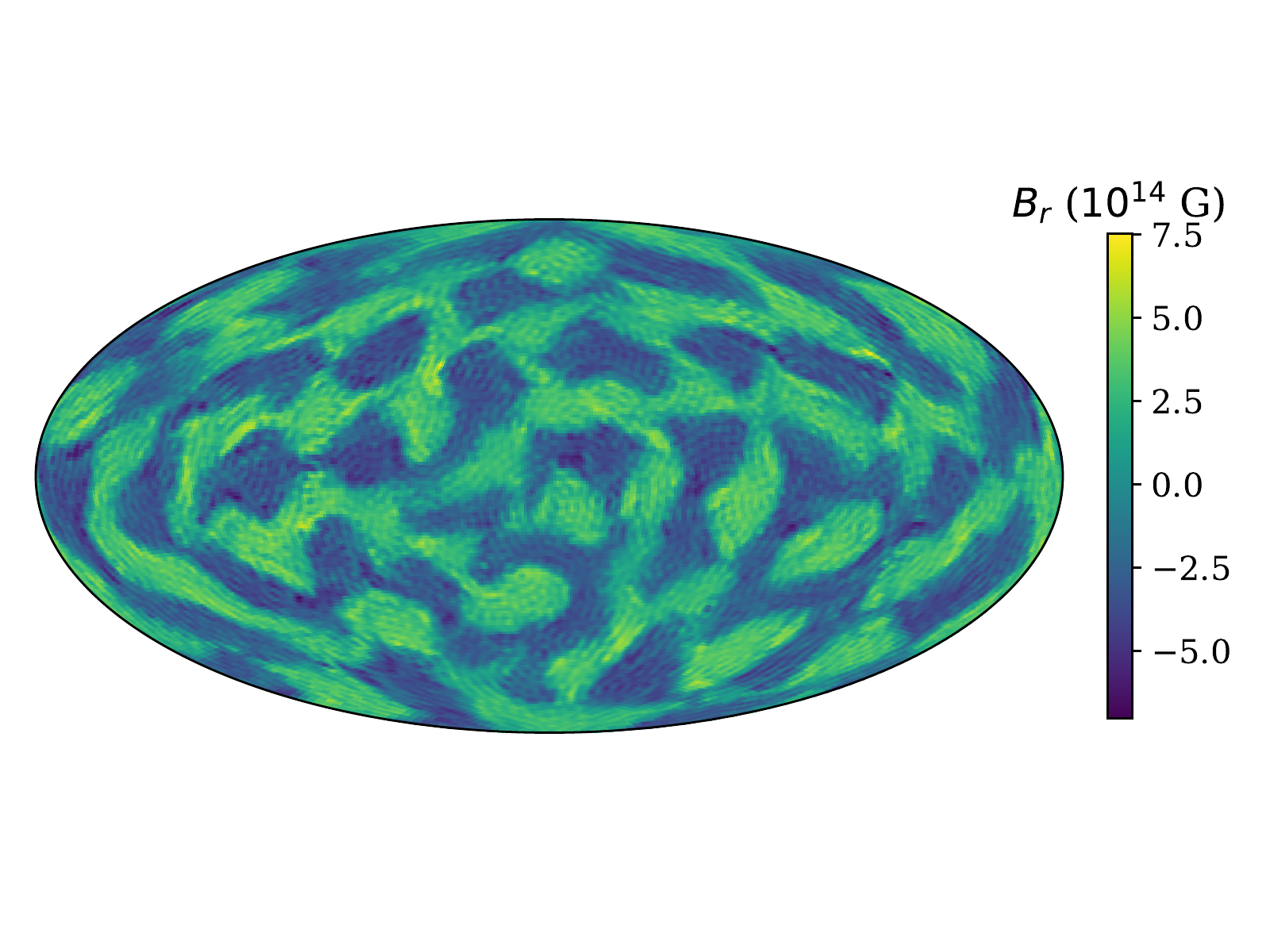}{0.3\textwidth}{(b)}
          \fig{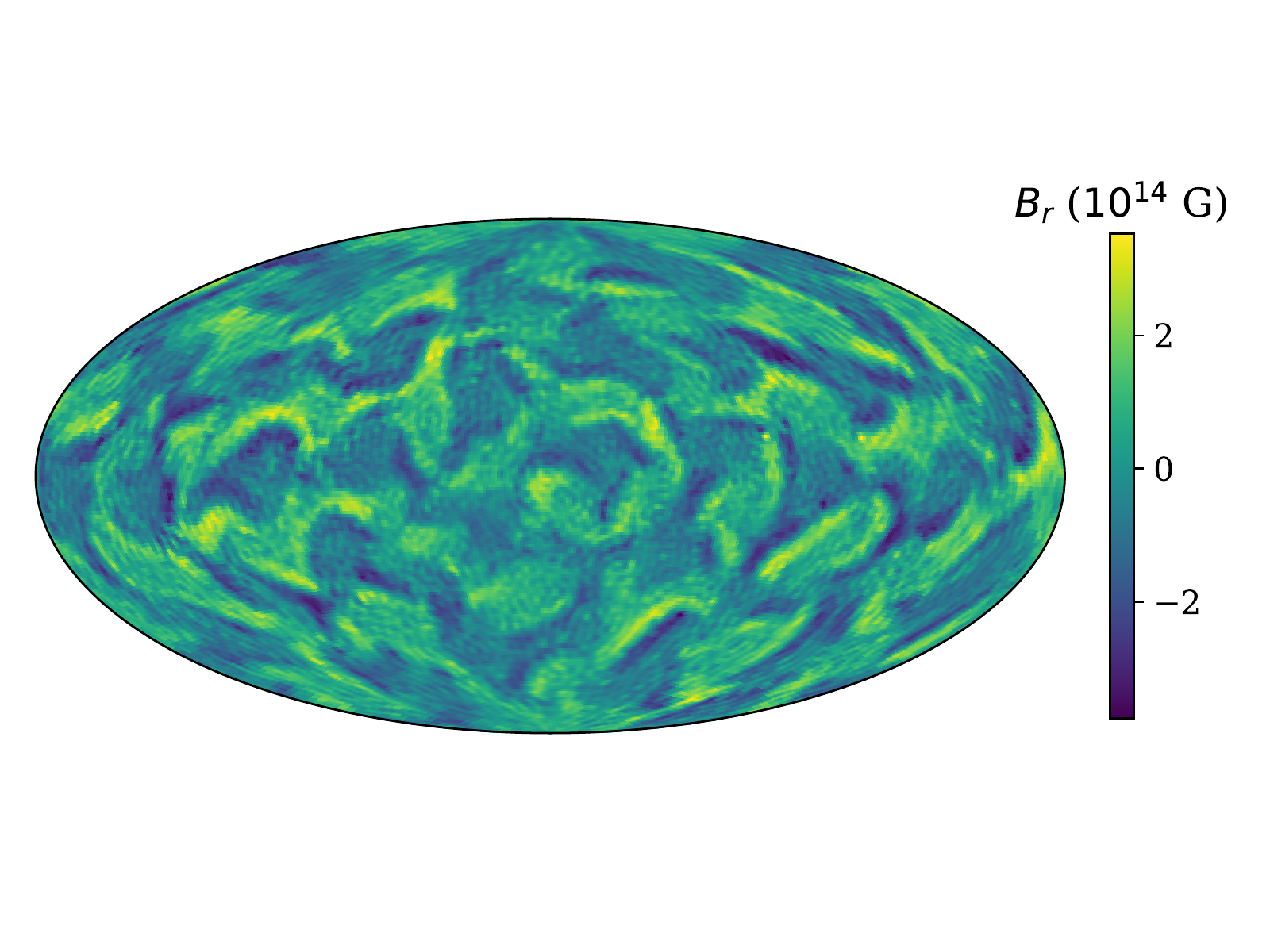}{0.3\textwidth}{(c)}
         }
\gridline{\fig{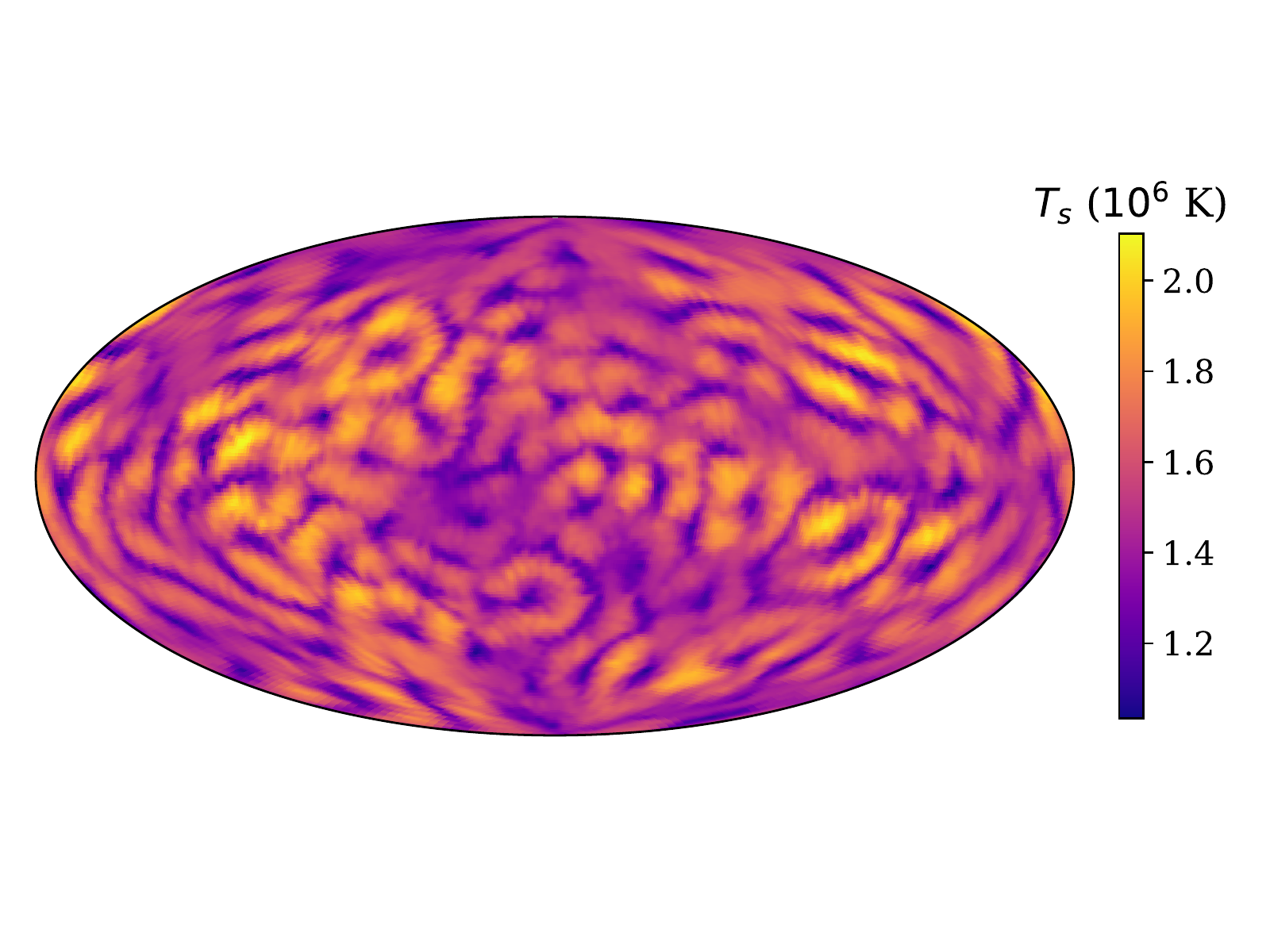}{0.3\textwidth}{(d)}
          \fig{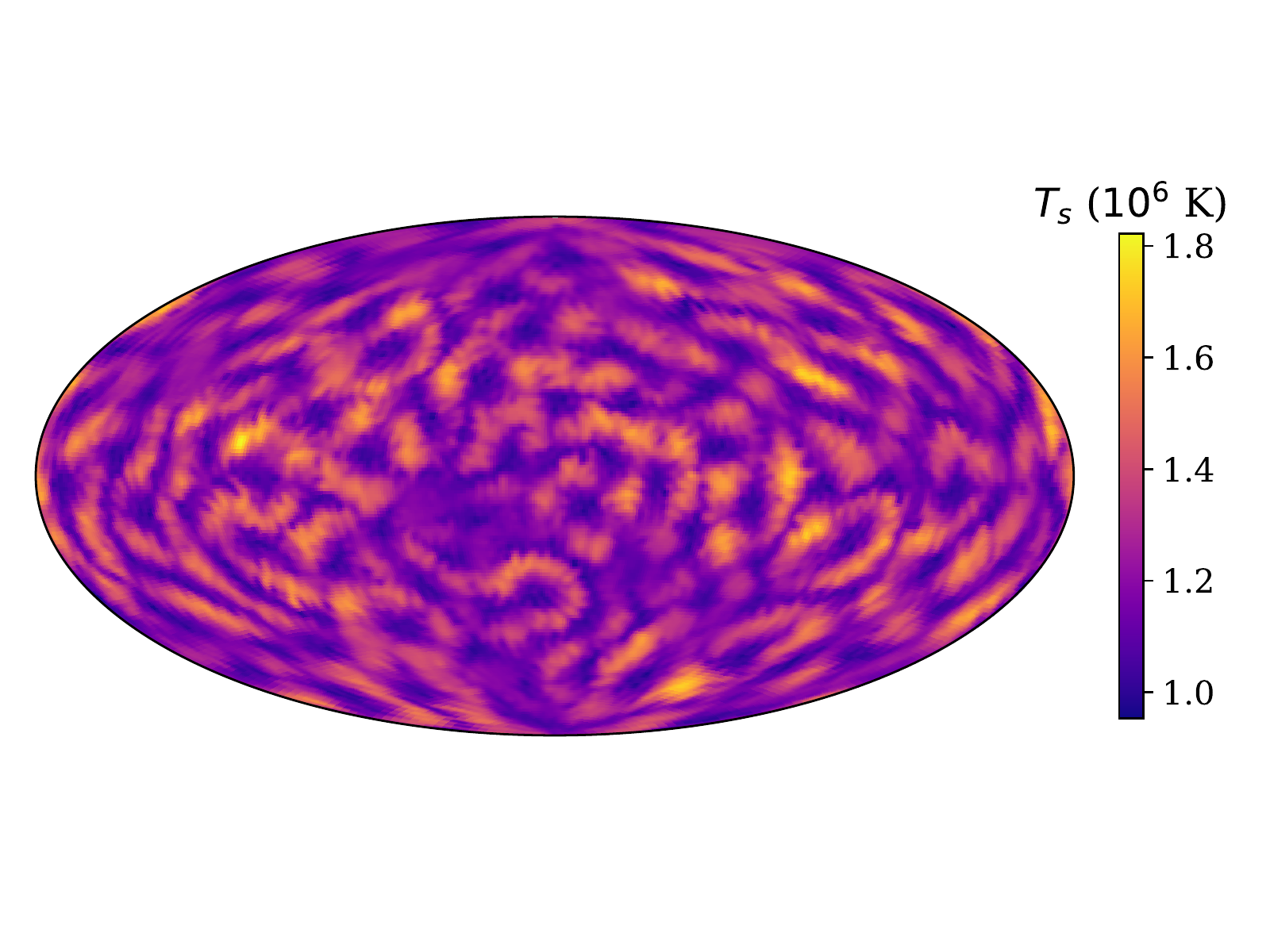}{0.3\textwidth}{(e)}
          \fig{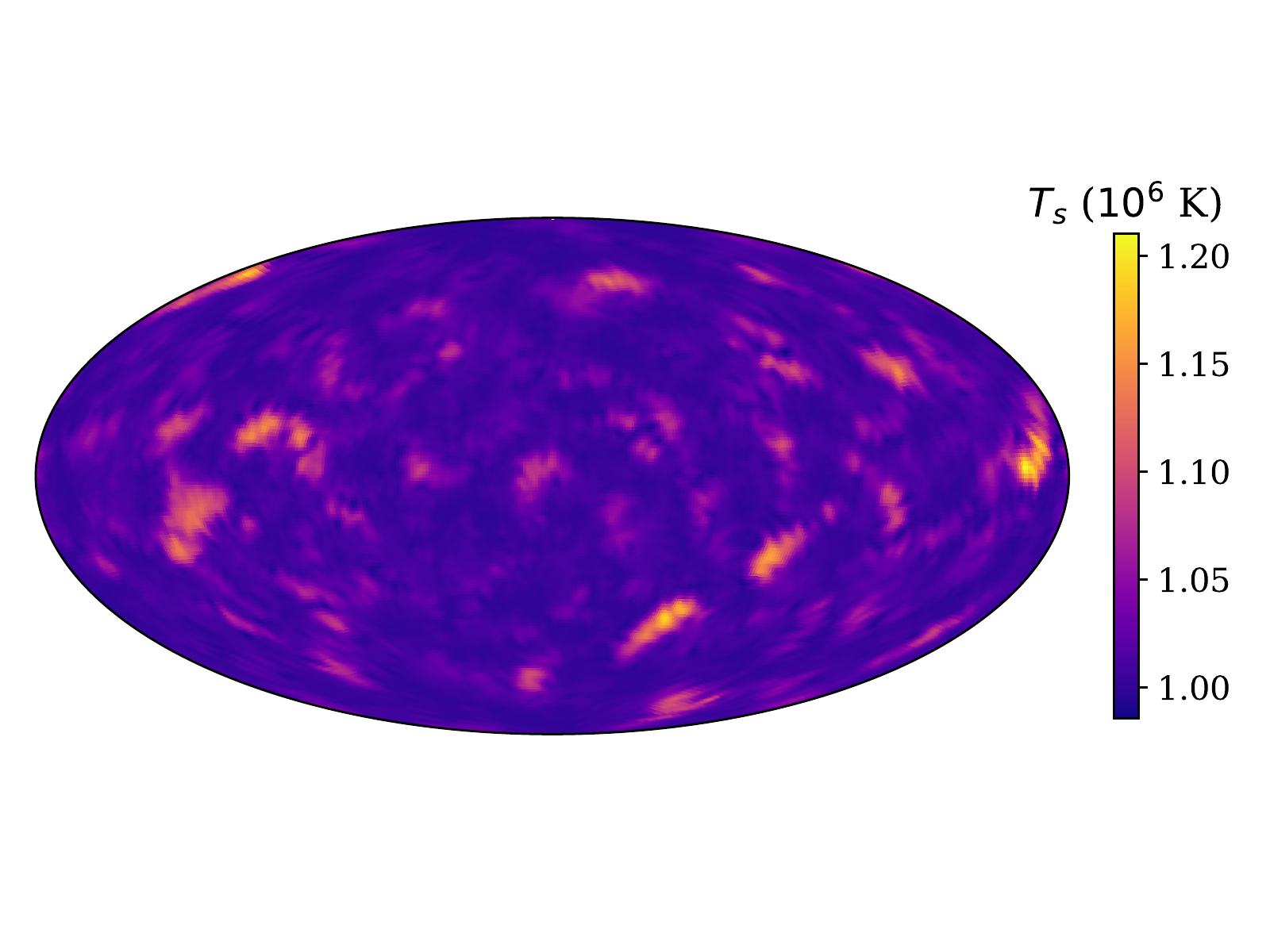}{0.3\textwidth}{(f)}
          }
\caption{Surface magnetic fields (upper row) and surface temperatures (lower row) for model 6. Individual panels correspond to different ages: (a, d) 3.5~Kyr, (b, e) 6.8~Kyr (c,f) 38~Kyr.
\label{f:m6}}
\end{figure*}

We computed models 1-3 up to 50~Kyr, and model 4,5,6 and 7 up to 100~Kyr. 
We show the surface temperature distributions for models 2, 4 and 6 in Figures~\ref{f:m2},~\ref{f:m4} and \ref{f:m6}. The surface temperature distributions in the case of model 1 and 3 are very similar to model 2. The surface temperatures of model 5 and 7 are very similar to ones produced in model 4 and 6, respectively, so we do not plot these maps. Basically the surface temperature distributions are defined only by the value of the initial magnetic energy, since the initial field topology is the same in all models. The initial strength of the dipolar component weakly affects the values of surface temperatures and the location of hot and cold regions.

In all cases the small-scale structure of the magnetic field leads to formation of complicated thermal pattern which includes alternating small hot and cold regions. 
Individual hot spots have simple shapes but temperatures of groups of spots differs in different parts of NS, making the pattern even more complicated.
Overall, at 3.5~Kyr the surface temperature map is composed of a few relatively small hot regions and hot spot associations, and a much more extended colder continuum.
We measure the linear sizes of several hot regions in the case of model 6 at 1.2~kyr. Two brightest hot regions have linear sizes of 2.2 and 2.6 km respectively, assuming that the NS radius is 10~km. At later stages of evolution ($>20$~kyr) the temperature and size of hot regions seen in all models decays. The hot regions become completely isolated and cool down. 

The hot regions are much hotter in the case of model 4 and 6 with larger initial magnetic energy than in model 2. In the case of model 6, which has the largest magnetic field energy, the hot spots are reaching temperatures of $2.16\times 10^6$~K, which is $\approx 0.19$~keV. \cite{gotthelf2013} mentioned that it is necessary for a successful CCO model to show variations of the temperature  by a factor of two over the surface, because this temperature difference is seen in observations of RX J0822-4300. Our models 4, 5, 6 and 7 show temperature variations with a factor of $1.7$~--~$2.2$ over the surface. 

\begin{figure}
    \centering
    \fig{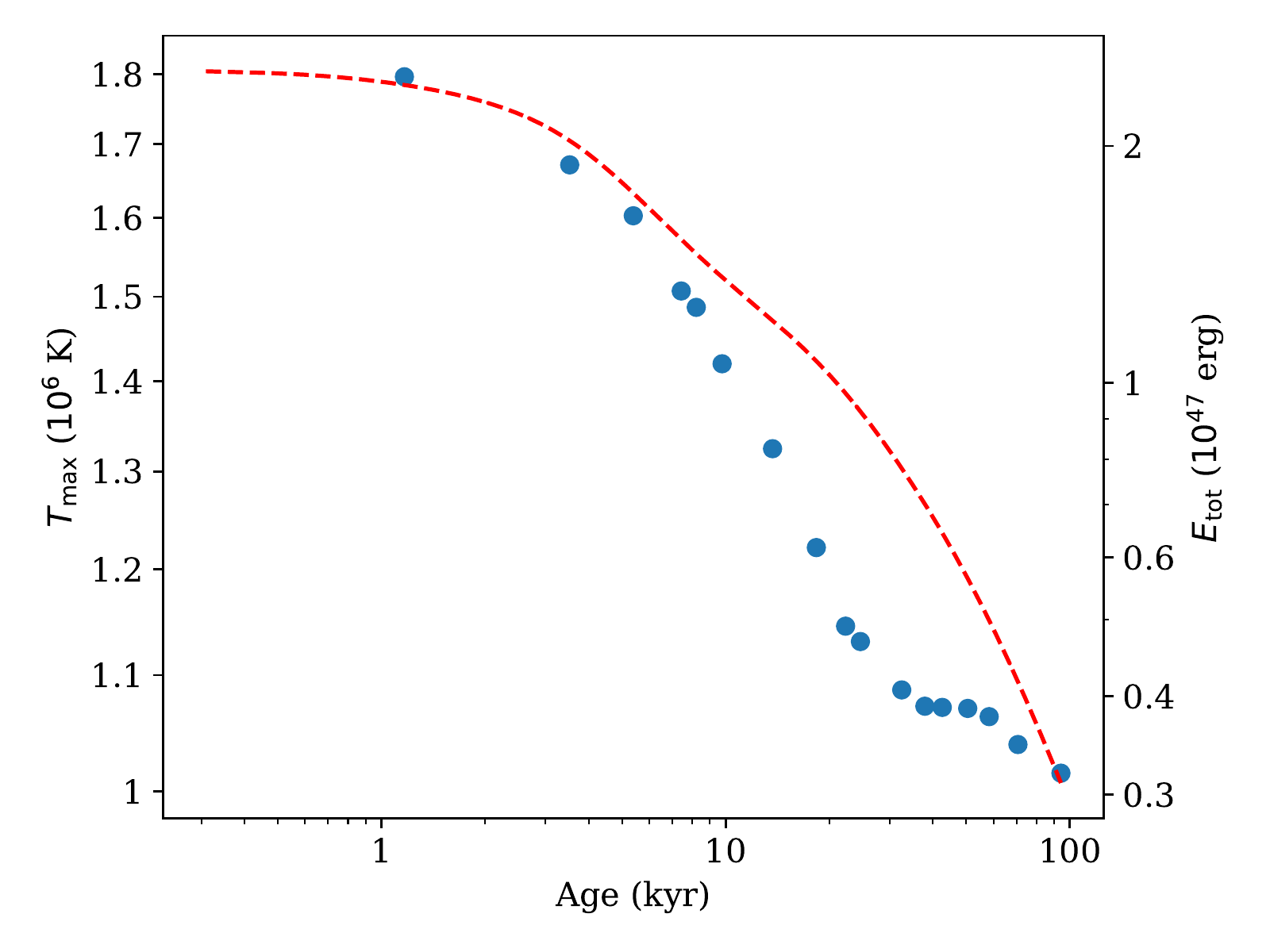}{0.45\textwidth}{}
    \caption{Evolution of maximum surface temperature (blue dots) and total magnetic energy (dashed red line) for model 4.}
    \label{f:temp}
\end{figure}

Surface heating is caused by the release of  magnetic energy stored in the crustal field, see Figure~\ref{f:temp}. During the first $\approx 20$~kyr, the initial total magnetic energy $E_\mathrm{tot}$ decays from $2.5\times 10^{47}$~erg to $\approx 5\times 10^{46}$~erg with typical power of $\approx 3\times 10^{35}$~erg/s. In real NSs this power is emitted through the surface photon emission and neutrinos. In our simulations this power is partly emitted through the surface boundary condition and partly absorbed by the inner boundary condition. 

\begin{figure*}
    \centering
    \gridline{\fig{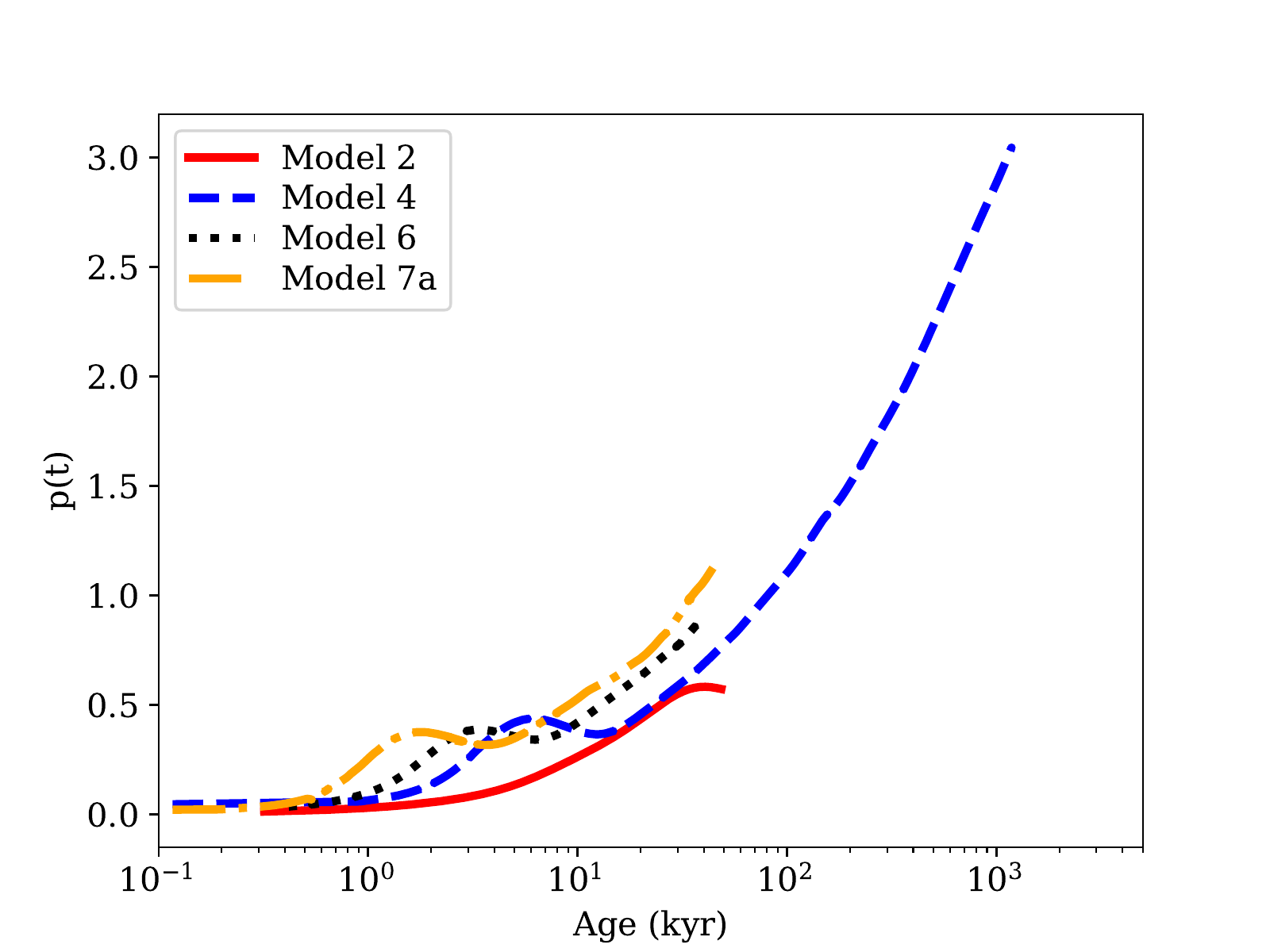}{0.48\textwidth}{(a)}
    \fig{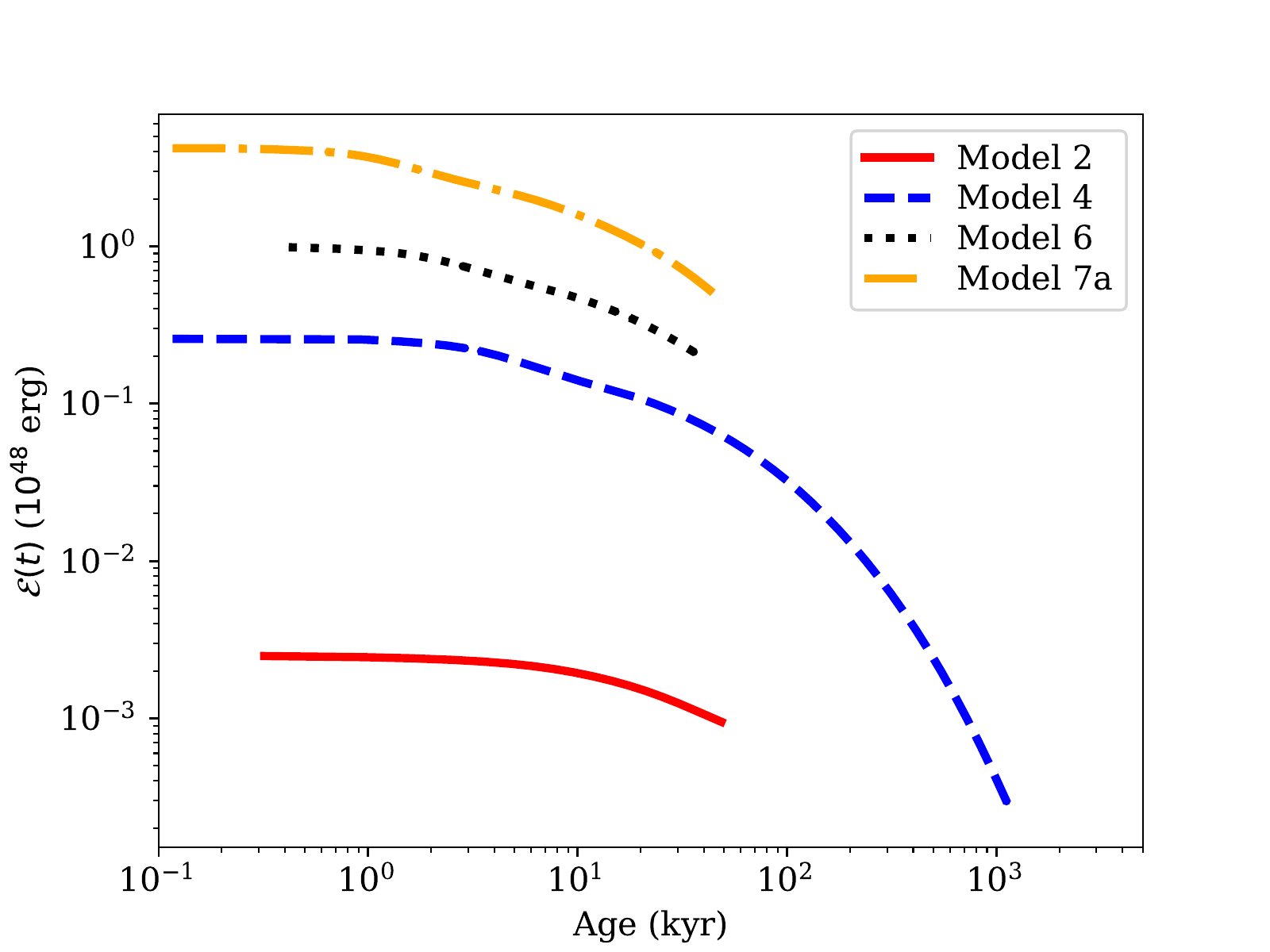}{0.48\textwidth}{(b)}
    }
    \caption{Evolution of the instantaneous energy exponent eq.~(\ref{energy}) (a) and total magnetic energy (b) for different models. Model 7a corresponds to model 7 in \cite{CCOI}.}
    \label{f:pt}
\end{figure*}

We estimate the instantaneous exponent which describes the decay of total magnetic energy:
\begin{equation}
\mathcal{E}(t) = \int E(k,t) dk     
\end{equation}
in a way similar to work by \cite{branbedburg2020}:
\begin{equation}
p(t) = -\frac{d\log \mathcal E (t)}{d\log t}
\label{energy}
\end{equation}
We show the results of this evolution in Figure~\ref{f:pt}. The behaviour of $p(t)$ is not monotonic. It reaches the first maximum with values ranging 0.39-0.58 (values for individual models can be found in Table~\ref{t:models}).
This level is reached at different physical times because the Hall timescale differs in these models. After this initial saturation, the $p(t)$ value briefly declines and starts growing again reaching values of 1-3 at the end of our simulations, see Figure~\ref{f:pt}. It means that by the end of our simulations the energy of magnetic field decays exponentially due to the Ohmic decay.  

The values of 0.39-0.58 corresponds to the value $2/5$ obtained for the case of helical magnetic fields by \cite{branbedburg2020}. He noticed that even magnetic fields with small fraction of helicity evolve quickly to nearly 100\% helical configurations. Therefore, it is not surprising that our simulations show an energy decay rate initially $\mathcal{E} = \mathcal{E}_0 t^{-2/5}$ which is typical for helical configurations.

\begin{figure}
    \centering
    \fig{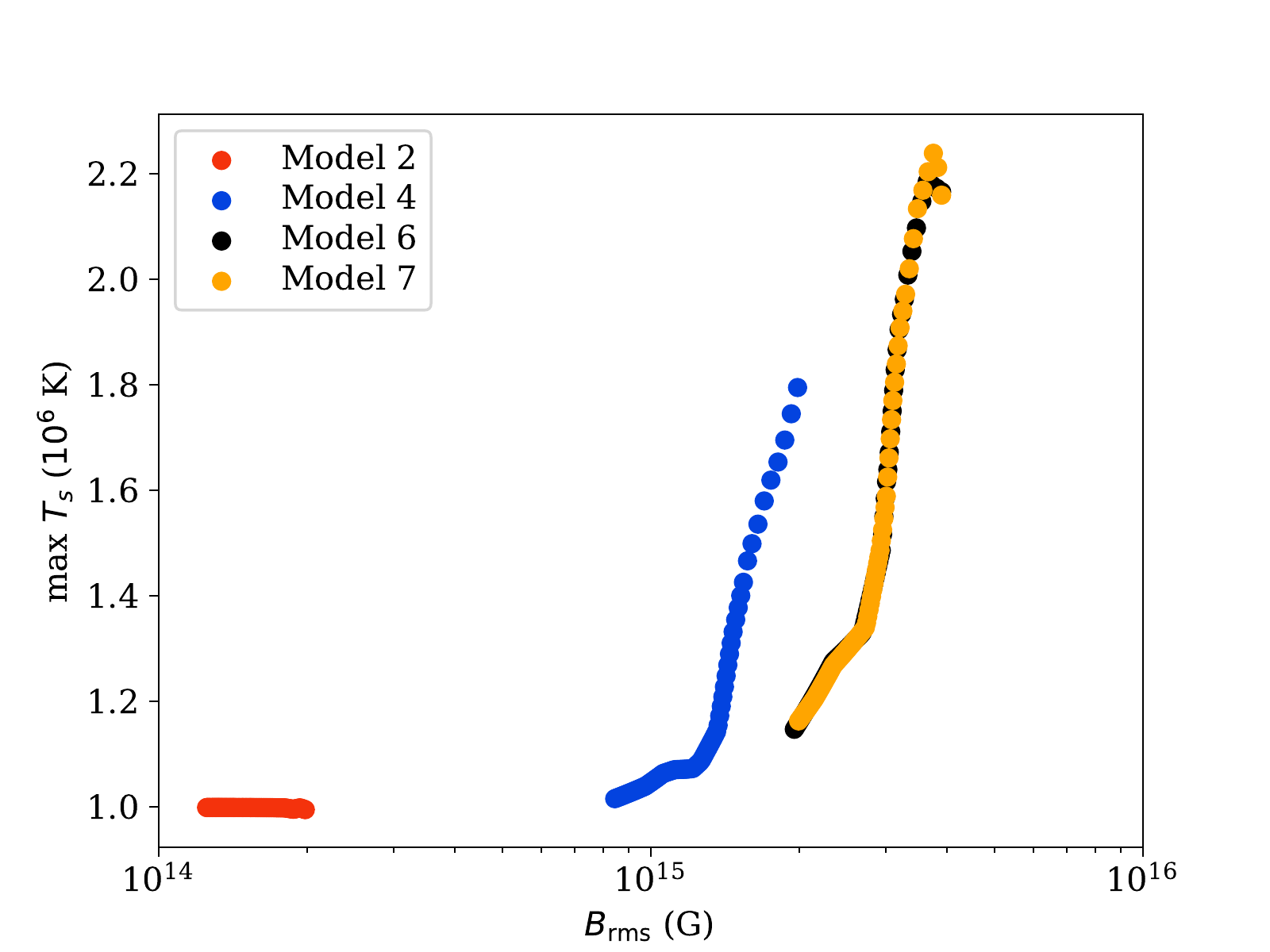}{0.45\textwidth}{}
    \caption{Evolution of maximum surface temperature as a function of $B_\mathrm{rms}$.}
    \label{f:brms_temp}
\end{figure}

\begin{figure}
    \centering
    \fig{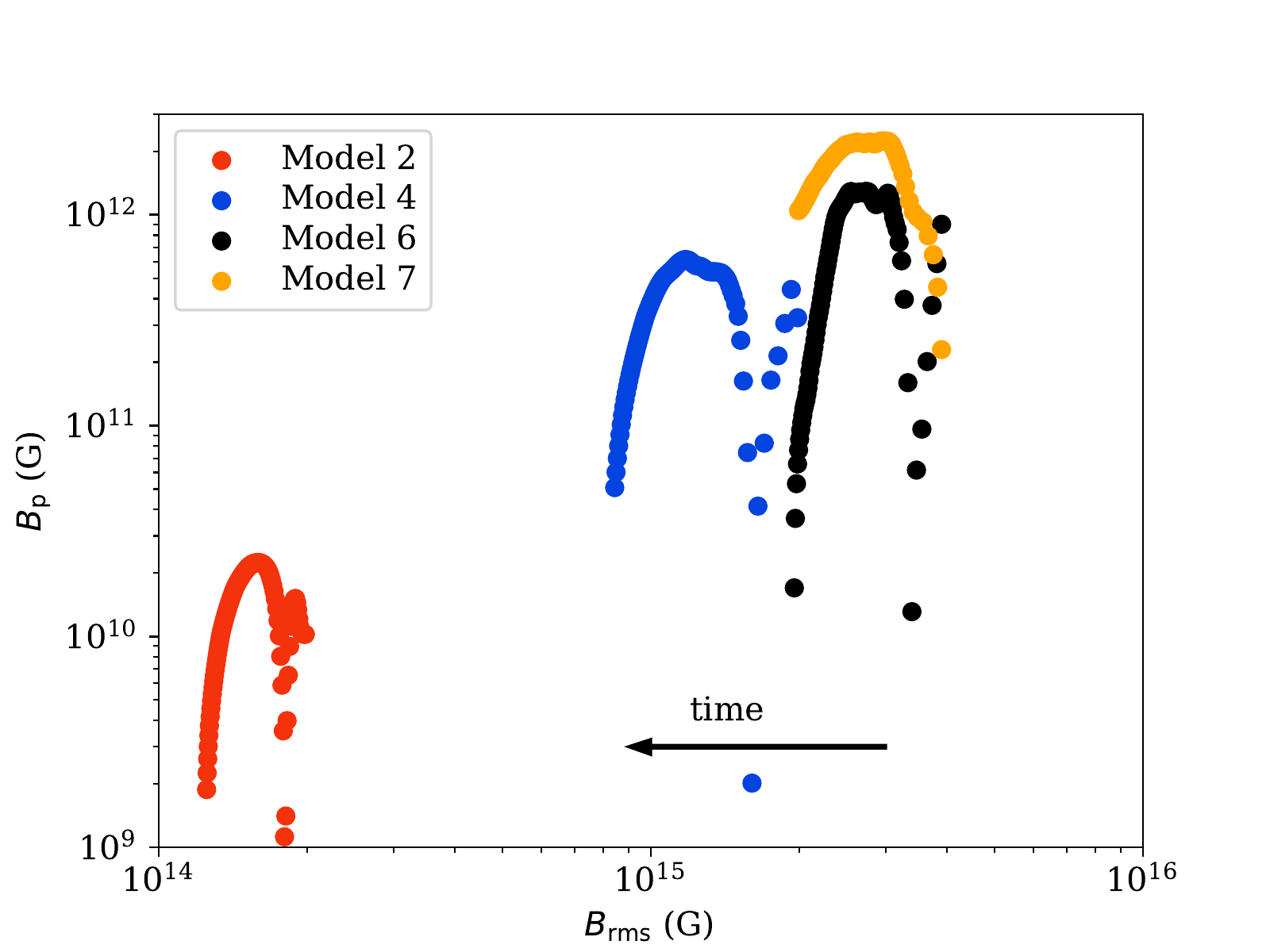}{0.45\textwidth}{}
    \caption{Evolution of dipolar poloidal component of the magnetic field as a function of $B_\mathrm{rms}$.}
    \label{f:brms}
\end{figure}

Another way to describe a complicated field using a single value is by using the root mean square of magnetic field $B_\mathrm{rms}$, similar to what \cite{branbedburg2020} obtained using the PENCIL code \citep{brandenburg2020b}. We compute $B_\mathrm{rms}$ over the whole volume of the NS crust. We show the evolution of the maximum temperature as a function of $B_\mathrm{rms}$ in Figure~\ref{f:brms_temp}. We see that similar values of $B_\mathrm{rms}$ might correspond to different maximum temperatures depending on the evolution stage.  On the other hand, the value of emerging dipolar, poloidal magnetic field seems to correlate with $B_\mathrm{rms}$, as seen in Figure~\ref{f:brms}. The initial value of the dipolar component gets erased quickly by the Hall evolution, and the field which emerges afterwards is related to $B_\mathrm{rms}$. In the case of model 2, the emergent field is $2\times 10^{10}$~G. In the case of models 6 and 7, the dipolar field reaches values of $\approx 10^{12}$~G, quite independently of the initial dipolar magnetic field. 

It is interesting to note that the Hall evolution creates vortices similar to turbulence seen in the surface temperature maps, especially in Figure~\ref{f:m2}, age 9.5~kyr. 
Typically the Hall evolution creates power-spectra somewhat similar to turbulence, but with a different slope of $l^{-2}$, see e.g.\ \cite{wareing2009,Goldreich1992}. When the Hall evolution starts with initial condition consisting of a strong global dipole, it mostly preserves the dipole component and redistributes a part of its energy into small-scale magnetic fields. The intensity of these fields is small in comparison to the dipole, and they are not distinguishable as spatial structures, see e.g.\ \cite{igoshev2020}. In simulation 2 we suppress the initial dipole and choose the small-scale harmonics to be strong at the beginning of the simulation. As the result we see that the magnetic energy is redistributed among small-scale harmonics in such a way that spatial vortices emerge in the thermal map,  which shows that under certain conditions the Hall turbulence is very similar to regular turbulence.
The system evolution of model 2 in scales larger than the crust scale-height $H$ is affected by the geometry of the system. This roughly corresponds to modes with  $\ell \approx \pi R_{NS}/H$,  with $\ell$ being the spherical harmonic decomposition mode and $R_{NS}$ the radius of the star. For a value of $H\sim 1$~km this corresponds to modes with $\ell>30$ where the behaviour of the turbulence will be more profound and the field will not be affected by the geometry of the system. 

\begin{figure*}
\gridline{\fig{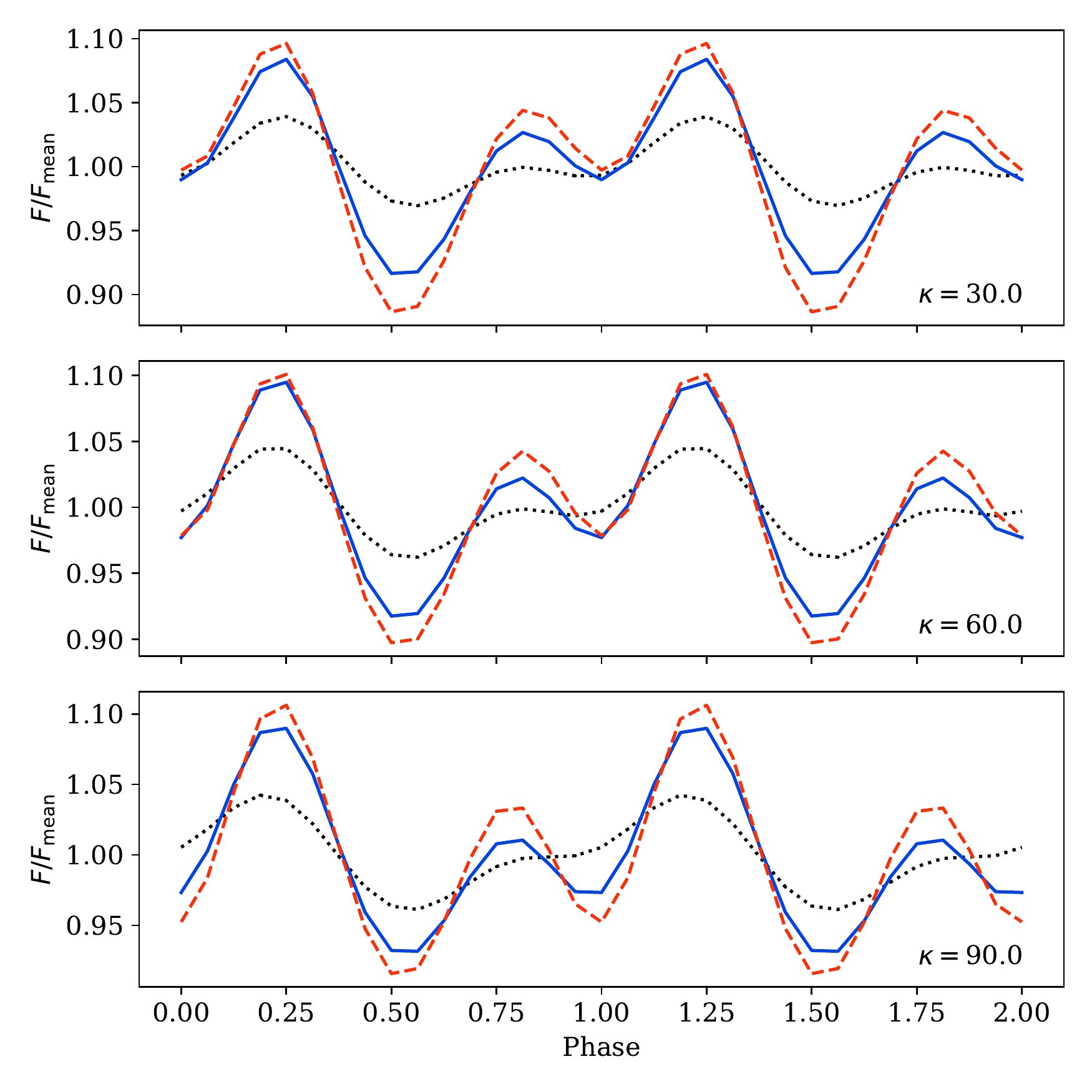}{0.48\textwidth}{(a)}
          \fig{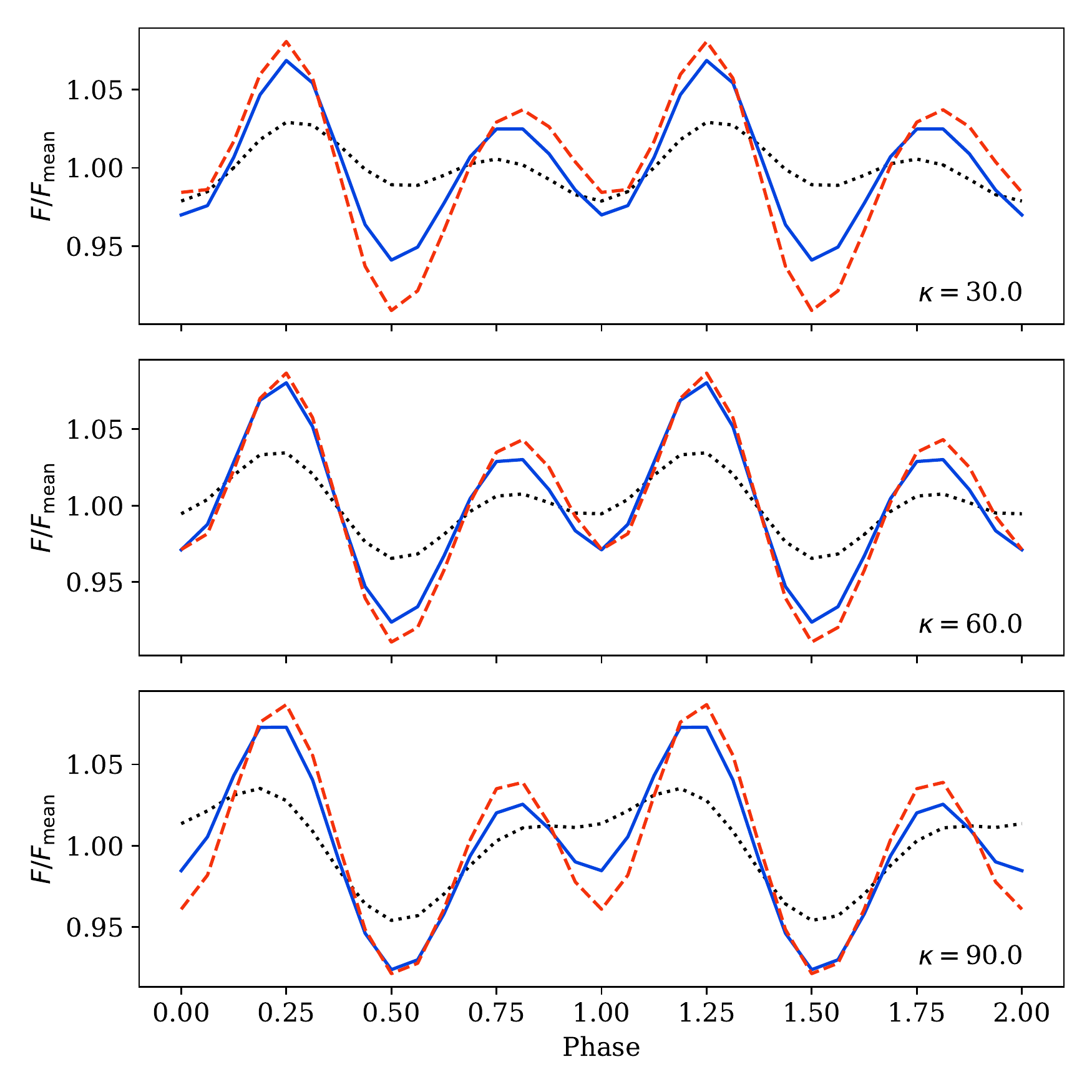}{0.48\textwidth}{(b)}
          }
\caption{Soft X-ray light curves for model 4 (left panel) and model 6 (right panel) at age 3.5~Kyr. Each panel corresponds to a different value of $\kappa$ (written in the lower right corner); black dotted line shows $i=30^\circ$, solid blue line corresponds to $i=60^\circ$ and red dashed line is for $i = 90^\circ$. 
\label{f:m4_lightcurve}}
\end{figure*}

Since the hot regions occur at multiple isolated locations at the surface, it means that a few of these regions are seen simultaneously, independently of the NS orientation. This effect is amplified further when the curved path of photons in the NS gravitational field is taken into account. We show the light curves for models 4 and 6 in Figure~\ref{f:m4_lightcurve}. Despite the factor two difference in temperature between hot and cold regions, the pulsed fraction is limited to 9-11\%, similar to real CCOs. 

Overall, the light curves have a few components which correspond to different hot regions. The shape of light curves is simple because the hot regions are compact. There is no use in discussing how much one light curve peak is higher/lower than another peak, because it is defined by the small-scale structure of the magnetic field. Slightly different configurations of the field could produce a very different light curve shape.

\section{Discussion}
\label{s:discuss}
The main caveats of our simulations are as follows: (1) we do not take NS neutrino cooling into account, (2) we use a simplified treatment of the NS atmosphere, (3) we assume a strong beaming of the thermal photon emission and (4) no magnetic field evolution in the core is assumed. Below we briefly explain how addition of these effects could change the results of our simulations.

NS cools down and the core temperature decays below $10^8$~K assumed in our simulations within first 10~kyr of the NS life. Therefore, the surface temperatures will be smaller in extended simulations. Nevertheless, the magnetic field isolates certain regions from the core and heats them up, so the temperature difference between hot and cold regions might increase in realistic simulations with NS neutrino core cooling. Another factor which is not taken into account is the neutrino cooling of the crust. It might remove the excess heat from hot spots, limiting their temperatures by some threshold.  

In our simulations we assume a temperature dependence between the deep crust and surface in form of eq.~(\ref{e:temp_s}). In reality the temperature depends on magnetic fields as well, see e.g.\ \cite{Potekhin2001}. This might change the surface temperature thermal map and could affect the light curves. We assume a strong beaming of the thermal emission proportional to $\cos^2 \theta'$, which approximately follows the exact numerical calculations of \cite{vanAdelsberg2006}. For weak magnetic fields the beaming might be weaker which will further decrease the maximum pulsed fraction which we see in our models.

Recently, \cite{Gusakov2020} modelled the ambipolar diffusion in the NS core. They noticed that magnetic field in the core could evolve significantly on timescales of NS cooling i.e.\ $10^4$~--~$10^6$ years \citep{igoshev2014,igoshev2015,igoshev2020mnras}. Such a magnetic field evolution proceeding at the lower boundary condition could noticeably change both the final magnetic field configuration arising as a result of magnetic field evolution in the crust and the surface thermal pattern.

\section{Conclusions}
\label{s:conclusion}

We perform three-dimensional magneto-thermal electron-MHD simulations of the NS crust to study the tangled magnetic field configurations. These configurations were suggested to be a possible mechanism to explain the peculiar emission properties of CCOs. 
We found that these magnetic field configurations lead to:
\begin{itemize}
    \item Formation of small hot regions with typical size of $\approx 2$~km which are located at significant separations from each other.
    \item The heating correlates with initial total magnetic energy. Most of the crustal heating due to the currents is released during the first 10-20 kyr of NS evolution. The power released due to the magnetic field decay could reach values of $3\times 10^{35}$~erg~s$^{-1}$. Only a fraction of this power is emitted as thermal X-ray radiation from the NS surface.
    \item In our simulations with initial total magnetic energy of $E_\mathrm{tot, 0} = 2.5\times 10^{47}$~erg, the hot regions have temperatures 1.7 times larger than the bulk surface temperature of NS. In our simulations with $E_\mathrm{tot, 0} = 10^{48}$~erg, the hot regions have temperatures which are 2.2 times larger than the bulk surface temperature. These factors are compatible with ones seen in the X-ray observations of CCOs.
    \item The maximum surface temperature decays exponentially on a timescale of $\approx 15$~kyr. The maximum temperature approaches the bulk surface temperature already after $\approx 20$~kyr. 
    \item The resulting light curves show modulations with maximum PFs of $9-11$~\% in the case of models 4, 5, 6 and 7. These small PFs can be explained by the fact that hot regions are located at multiple positions on the NS, and it is impossible to choose such an orientation where none are present. This is a reason why larger $E_\mathrm{tot,0}$ does not necessarily lead to an increase in PF. Regions become hotter, but a few of them are still seen simultaneously.
    \item The final poloidal magnetic field correlates with root mean square magnetic field. 
\end{itemize}

Overall, the hidden magnetic field could provide enough energy to explain enhanced thermal emission of CCOs. A part of the tangled magnetic field energy is released through thermal emission from NS surface via emission of small hot spots. The small size of these spots is related to a typical size of magnetic field loops assumed as the initial condition in our simulations. These sizes are comparable to the NS crust depth and could be produced in a stochastic dynamo. Therefore, the main difference between CCOs and magnetars in this scenario is the typical size of the magnetic field. In magnetars the magnetic field is large-scale, while in CCOs it is mostly small-scale.

\section*{Acknowledgements}
This work was supported by STFC grant ST/S000275/1. 
This work was undertaken on ARC3 and ARC4, part of the High Performance Computing facilities at the University of Leeds, UK.




\bibliography{sample63}{}
\bibliographystyle{aasjournal}



\end{document}